\documentclass[useAMS,usenatbib]{./aa}
\usepackage{txfonts}
\usepackage{natbib}
\bibpunct{(}{)}{;}{a}{}{,}
\usepackage[dvips]{graphicx}
\begin{document}
\title{X-ray properties in massive galaxy clusters: XMM-Newton observations of the REFLEX-DXL sample
\thanks{This work is based on observations
made with the XMM-Newton, an ESA science mission with
instruments and contributions directly funded by
ESA member states and the USA (NASA).}}
\author{Y.-Y. Zhang\inst{1},
H. B\"ohringer\inst{1},
A. Finoguenov\inst{1},
Y. Ikebe\inst{1,2},
K. Matsushita\inst{1,3},
P. Schuecker\inst{1},
L. Guzzo\inst{4} and
C. A. Collins\inst{5}}
\institute{Max-Planck-Institut f\"ur extraterrestrische Physik, Giessenbachstra\ss e, 85748 Garching, Germany
\and National Museum of Emerging Science and Innovation, Tokyo, Japan
\and Tokyo University of Science, Tokyo, Japan
\and INAF - Osservatorio Astronomico di Brera, Merate/Milano, Italy
\and Liverpool John Moores University, Liverpool, U.K.}
\authorrunning{Zhang et al.}
\titlerunning{X-ray properties in massive galaxy clusters}
\date{Received 17 June 2005 / accepted 5 June 2006}

\offprints{Y.-Y. Zhang}

\abstract{We selected an unbiased, flux-limited and almost
volume-complete sample of 13 distant, X-ray luminous (DXL, $z\sim
0.3$) clusters and one supplementary cluster at $z=0.2578$ from the
REFLEX Survey (the REFLEX-DXL sample). We performed a detailed study
to explore their X-ray properties using XMM-Newton observations. Based
on the precise radial distributions of the gas density and
temperature, we obtained robust cluster masses and gas mass
fractions. The average gas mass fraction of the REFLEX-DXL sample at
$r_{500}$, $0.116 \pm 0.007$, agrees with the previous cluster studies
and the WMAP baryon fraction measurement. The scaled profiles of the
surface brightness, temperature, entropy, gas mass and total mass are
characterized by a self-similar behaviour at radii above
0.2--0.3~$r_{500}$. The REFLEX-DXL sample confirms the previous
studies of the normalization of the scaling relations ($L$--$T$,
$L$--$M$, $M$--$T$ and $M_{\rm gas}$--$T$) when the redshift evolution
of the scaling relations is accounted for. We investigated the scatter
of the scaling relations of the REFLEX-DXL sample. This gives the
correlative scatter of (0.20,0.10) for variable of ($M$,$T$) of the
$M_{500}$--$T$ relation, for example.

}

\maketitle

\section{Introduction}

The number density of galaxy clusters probes the cosmic evolution of
large-scale structure (LSS) and thus provides an effective test of
cosmological models.  It is sensitive to the matter density,
$\Omega_{\rm m}$, and the amplitude of the cosmic power spectrum on
cluster scales, $\sigma_8$ (e.g. Schuecker et al. 2003).  Its
evolution is sensitive to the dark energy, the density of which is
characterized by the parameter $\Omega_{\Lambda}$ and the equation of
state parameter $w(z)$ (e.g. Vikhlinin et al. 2002; Allen et al. 2004;
Chen \& Ratra 2004).

The most massive clusters show the strongest and cleanest effects in
the cosmological evolution. The structure of the X-ray emitting
intra-cluster plasma in massive galaxy clusters is predominantly
determined by gravitational effects and shock heating and is less
affected by non-gravitational processes than low mass clusters. Only
with decreasing cluster mass and intra-cluster medium (ICM)
temperature, non-gravitational effects play an important role before
and after the shock heating (Voit \& Bryan 2001; Voit et al. 2002;
Zhang \& Wu 2003; Ponman et al. 2003). The X-ray properties of the
most massive clusters are thus well described in hierarchical
modeling. Therefore, massive galaxy clusters are especially important
in tracing LSS evolution. The most massive clusters provide the
cleanest results in comparing theory with observations.

Excluding cooling cores (Fabian \& Nulsen 1977), a self-similar
behaviour of the distributions of the ICM properties such as the
temperature, density and entropy of massive clusters ($>$~4~keV)
is indicated in ROSAT, ASCA, Chandra and XMM-Newton observations
(e.g. Markevitch 1998; Markevitch et al. 1998; Vikhlinin et al.
1999, 2005, 2006; Arnaud et al. 2002; Reiprich \& B\"ohringer 2002;
Zhang et al. 2004a, 2004b, 2005b; Ota \& Mitsuda 2005; Pratt \&
Arnaud 2005; Pointecouteau et al. 2005) and simulations (e.g.
Borgani 2004; Borgani et al. 2004; Kay 2004; Kay et al. 2004). As
a consequence of the similarity, massive galaxy clusters show
tight scaling relations such as the luminosity--temperature
($L$--$T$, e.g. Isobe et al. 1990; Markevitch 1998; Arnaud \&
Evrard 1999; Ikebe et al. 2002; Reiprich \& B\"ohringer 2002),
luminosity--mass ($L$--$M$, e.g. Reiprich \& B\"ohringer 2002;
Popesso et al. 2005), mass--temperature ($M$--$T$, e.g. Nevalainen
et al. 2000; Finoguenov et al. 2001b; Neumann \& Arnaud 2001; Xu
et al. 2001; Horner 2001; Reiprich \& B\"ohringer 2002; Sanderson
et al. 2003; Pierpaoli et al. 2001, 2003), and
luminosity--metallicity ($L$--$Z$, e.g. Garnett 2002) relations.
Additionally, a reliable estimate of the intrinsic scatter of the
scaling relations is the key to a correct modeling to constrain
cosmological parameters in the cosmological applications using
galaxy clusters. Therefore, understanding the intrinsic scatter of
the scaling relations is as important as studying the scaling
relations themselves. For example, on-going cluster mergers
partially account for the scatter in the scaling relations since
on-going mergers may lead to a temporary increase not only in the
temperature and X-ray luminosity (Randall et al. 2002), but also
in the core radius and slope of the surface brightness profile.
Precise measurements of the ICM structure are required to allow
accurate cluster mass and gas mass fraction determinations and
thus to investigate the X-ray scaling relations for comparison of
theory with simulations and observations.

The ROSAT-ESO Flux-Limited X-ray (REFLEX, B\"ohringer et al. 2001,
2004) galaxy cluster survey provides the largest homogeneously
selected catalog of X-ray clusters of galaxies so far. It provides
the basis to construct an unbiased sub-sample of clusters with
specific selection criteria. We exploit it to compose a sample of
distant, X-ray luminous (DXL) galaxy clusters in the redshift
range, $z=0.27$ to $0.31$, with $L_{X}~\geq~10^{45}~{\rm
erg~s^{-1}}$ for the 0.1--2.4~keV band and one supplementary
cluster at $z=0.2578$ (the REFLEX-DXL sample)\footnote{An
Einstein-de Sitter cosmological model with $\Omega_{\rm m} = 1.0$,
$\Omega_{\Lambda} = 0.0$ and the Hubble constant $H_{\rm
0}=50$~km~s$^{-1}$~Mpc$^{-1}$ was used for the $L_{X}$ threshold
in the sample construction. This luminosity threshold corresponds
to $L_{X}~\geq~5.9 \times 10^{44}~{\rm erg~s^{-1}}$ for a flat
$\Lambda$ cold dark matter ($\Lambda$CDM) cosmology with the
density parameter $\Omega_{\rm m}=0.3$ and the Hubble constant
$H_{\rm 0}=70$~km~s$^{-1}$~Mpc$^{-1}$}. The volume completeness
correction can be done using the well known selection function of
the REFLEX survey (B\"ohringer et al. 2004).

Prime goals for the study of the REFLEX-DXL sample are, (1) to obtain
reliable ICM properties such as temperature structure (Zhang et
al. 2004a, Paper\,II), (2) to determine accurate cluster masses and
gas mass fractions using precise ICM property measurements, (3) to
measure the normalization of the scaling relations with an improved
accuracy and to discuss their intrinsic scatter in detail, and (4) to
test the evolution of both the scaling relations and the temperature
function (e.g. Henry 2004) by comparing the REFLEX-DXL sample (at $z
\sim 0.3$) to nearby cluster samples. We present the results from high
quality XMM-Newton data in this work. The data reduction is described
in Sect.~\ref{s:method}. We derive the X-ray properties of the ICM and
determine the total masses and gas mass fractions based on precise gas
density and temperature radial profiles in Sect.~\ref{s:result}). We
investigate the self-similarity of the REFLEX-DXL clusters
(Sect.~\ref{s:scale}) and discuss the peculiarities of the individual
clusters accounting for the scatter around the self-similarity
(Sect.~\ref{s:discu}). In Sect.~\ref{s:conclusion}, we draw
conclusions. Unless explicitly stated otherwise, we adopt a flat
$\Lambda$CDM cosmology with the density parameter $\Omega_{\rm m}=0.3$
and the Hubble constant $H_{\rm 0}=70$~km~s$^{-1}$~Mpc$^{-1}$. We
adopt the solar abundance values of Anders \& Grevesse
(1989). Confidence intervals correspond to the 68\% confidence
level. We apply the Orthogonal Distance Regression method (ODR;
e.g. Feigelson \& Babu 1992; Akritas \& Bershady 1996) and take into
account measurement errors on both variables for the parameter fitting
of the scaling relations. We use Monte Carlo simulations for the
uncertainty propagation on all quantities of interest.

\section{Data reduction}
\label{s:method}

\subsection{Data preparation}

All 14 REFLEX-DXL clusters were observed by XMM-Newton in AO-1. Some
properties of these observations and an overview of the sample are
described in B\"ohringer et al. (2006, Paper\,I). The observations of
5 clusters in AO-1 were heavily contaminated by flares. Zhang et
al. (2004a) investigated the temperature structure of 9 remaining
REFLEX-DXL clusters observed in AO-1. Finoguenov et al. (2005) applied
a 2-dimensional approach to study the structure such as projected
density, temperature, pressure and entropy maps for those 9 REFLEX-DXL
clusters. All 5 clusters flared in AO-1 were re-observed in AO-3 in
which four have sufficient quality for a detailed study. All the
clusters of the REFLEX-DXL sample were uniformly analyzed in this
work.

All observations were performed with thin filter for three
detectors. The MOS data were taken in Full Frame (FF) mode. The pn
data were taken in Extended Full Frame (EFF) mode in AO-1 and FF mode
in AO-3, respectively. For pn, the fractions of the out-of-time (OOT)
effect are 2.32\% and 6.30\% for the EFF mode and FF mode,
respectively. An OOT event file is created and used to statistically
remove the OOT effect.

Good calibration is required for correct data reduction. The
difference of the spectral measurements using low energy cut-off
values of 1~keV and 0.4~keV was significant at the 1-$\sigma$
confidence level for most data sets in the previous analysis using the
calibration in XMMSAS v5.4.0 (Zhang et al. 2004a). Now we use the
XMMSAS v6.5.0 software for the data reduction. The new calibration in
XMMSAS v6.5.0 provides a better agreement between data and model for
EPIC, which are described in XMM-SOC-CAL-TN-0018 (Kirsch 2005). The
CTI correction, point-spread function (PSF) core, astrometry, gain and
energy redistribution have been improved. For example, in XMMSAS v6.0
the photon energy redistribution was strongly increased for energies
below 0.53~keV in order to reduce excesses seen at low
energies. Subsequent analysis of EPIC-pn spectra revealed large
residuals around 0.43~keV which clearly showed that the redistribution
was wrongly modeled in XMMSAS v6.0. The values in the redistribution
matrix at low energies using the ground calibration were replaced
based on the analysis of a set of spectra in the 0.1--2~keV band. For
XMMSAS v6.5.0 the redistribution is performed energy-dependently to
flattening the residuals. The spectra below 0.5~keV are greatly
improved (Kirsch 2005). The spectral measurements using low energy
cut-off values of 1~keV and 0.4~keV are now consistent for most
REFLEX-DXL clusters except for RXCJ0658$-$5556 ($\sim $~15\%). With
the new calibration in XMMSAS v6.5.0, we therefore use the 0.4--10~keV
band for the spectral analysis.

Above 10~keV (12~keV), there is little X-ray emission from clusters
for MOS (pn) due to the low telescope efficiency at these
energies. The particle background therefore dominates. The light curve
in the range 10--12~keV (12--14~keV) for MOS (pn), binned in 100~s
intervals, is used to monitor the particle background and to excise
periods of high particle flux. Since episodes of ``soft proton
flares'' (De Luca \& Molendi 2004) were detected in the soft band, the
light curve of the 0.3--10~keV band, binned in 10~s intervals, is used
to monitor and to excise the soft proton flares. A 10~s interval bin
size is chosen for the soft band to provide a similar good photon
statistic as for the hard band. The average and variance of the count
rate (ctr) have been interactively determined for each light curve
from the ctr histogram. Good time intervals (GTIs) are those intervals
with ctrs below the threshold, which is defined as 3-$\sigma$ above
the average. The GTIs of both the hard band and the soft band are used
to screen the data. The background observations are screened by the
GTIs of the background data, which are produced using exactly the same
thresholds as for the corresponding cluster target field. All the
observations of RXCJ2011.3$-$5725 were almost completely contaminated
by flares. In total, 183 pn counts (0.4--10~keV) are available for the
imaging spectral analysis of this cluster. RXCJ2011.3$-$5725 is
therefore excluded in the detailed study. Settings of $FLAG=0$ and
$PATTERN<13$ ($PATTERN<5$) for MOS (pn) are used in the screening
process.

\subsection{Source detection}

Over half of the clusters show clear substructures or/and
elongation. An ``edetect\_chain'' command has been used to detect
point-like sources. Point sources except for those suspicious
candidates described below are subtracted before the further data
reduction.

In the cluster center, point-like sources are hard to identify because
their emission is blended with the strongly peaked cluster emission,
in particular in cooling core clusters (CCCs). At the angular
resolution of the observations, it is difficult to distinguish between
a steep cooling core cusp and a central point source (e.g. an AGN in a
central dominant galaxy). In principle, cluster emission is thermal
and point source emission is non-thermal. The spectral shape can be
used to identify point sources because a power law shape usually
indicates the non-thermal emission contributed by point sources. We
studied the spectra of the regions where there is a possible presence
of point source in the cluster centers. We found that both ``mekal''
and ``powerlaw'' models provide acceptable fits. The photon statistic
does not allow for a clear discrimination of the spectral
shape. Therefore, a contribution of point sources to the cooling core
cusp can not be ruled out for these clusters. The properties of these
central regions are listed in Table ~\ref{t:centpsrc}. However, those
suspicious candidates are not subtracted before the further data
reduction.

\subsection{Background subtraction}

The background consists of several components exhibiting different
spectral and temporal characteristics (e.g. De Luca \& Molendi
2001; Lumb et al. 2002; Read \& Ponman 2003). The background
components can approximately be divided into two groups (e.g.
Zhang et al. 2004a). Group I contains the background components
showing significant vignetting (hereafter BVIGs), e.g. the cosmic
X-ray background (CXB). Group II contains the components showing
little or no vignetting (hereafter BNVIGs), e.g. particle-induced
background.

It is safe to use a local background only if the vignetting effect is
similar for the source and background regions. It becomes less
reliable when large vignetting corrections are needed. The vignetting
effect is energy-dependent and becomes significant above
5~keV. Additionally, the background component caused by the
instrumental lines is different from position to position. This cannot
be accounted for by a local background. The REFLEX-DXL cluster
emission covers most of the field of view (FOV), $r \sim
8^{\prime}$. A local background is only available in the outskirts
(e.g. $9.2^{\prime}<r<11.5^{\prime}$) but requires substantial
vignetting corrections. When the local background is corrected
for vignetting effects, the particle component is over corrected
especially at high energies (artificially increased). This leads to an
artificial softening of the derived cluster spectrum. The
temperatures of the REFLEX-DXL clusters are above 5~keV, and thus tend
to show a soft excess when the local background is used. Therefore, a
local background is not the best choice for the analysis of the
REFLEX-DXL sample.

Suitable background observations for such hot clusters are
XMM-Newton observations of almost blank fields using the same
instrumental set-up (e.g. a particular filter). We thus choose the
blank sky accumulations in the Chandra Deep Field South (CDFS) as
background. The CDFS observations used the same filter (thin
filter for all detectors) and mode (FF/EFF mode for MOS/pn) as the
REFLEX-DXL observations. In Zhang et al. (2005a), we investigated
the blank sky accumulations in Lumb et al. (2002) as an
alternative background and compared it to the CDFS observations.
We found that the measurements are consistent within 1-$\sigma$ in
both the spectral and the surface brightness analyses.

Suitable background observations guarantee similar BVIGs as for the
targets in the same detector coordinates. One can subtract such a
background extracted in the same detector coordinates as for the
targets. However, the difference between the target and background
should be taken into account as a residual background in the
background subtraction. This can be done for the REFLEX-DXL clusters
because the cluster X-ray emission does not cover the whole FOV, but
the $r<9^{\prime}$ region. The outskirts in the FOV (e.g.
$9.2^{\prime}<r<11.5^{\prime}$) were used to analyze the residual
background. Strictly, the residual background, obtained after
subtracting the blank field data, consists of both the residual CXB
(because it varies on the sky) and the particle induced background
(e.g residual background induced by soft protons).

\subsection{Substructure}

As an unbiased sample with respect to cluster morphology, the
REFLEX-DXL clusters are characterized in terms of their dynamical
state (Table~\ref{t:dyn}, also see Zhang et al. 2006) using the
classification in Jones \& Forman (1992). The substructure
consideration is based on the cluster morphology classification. The
detection of the cluster morphology and substructures strongly depends
on the cluster brightness, spatial resolution, and exposure time
(Jones \& Forman 1992). We define the substructure regions using the
0.5--2~keV image as follows.

As a start, we cut out the regions obviously contaminated by
substructures. The remaining region can be fitted by a symmetric
surface brightness model convolved with the PSF plus the scaled CDFS
background. We simulated a 0.5--2~keV image by convolving the
symmetric surface brightness model with the PSF plus the scaled CDFS
background. We subtracted the simulated image from the observed
0.5--2~keV image and obtained a residual image. The summed counts of
the substructure region in the residual image characterize the
substructure signal ($S$). The summed counts of the same region in the
simulated image are used as the background ($B$). We scaled the
residual image using the simulated image and defined a series of
contour regions. The signal-to-noise-ratio for each contour region was
calculated by $S/N=S/ \sqrt {S^2+B^2}$. The region corresponding to
the $S/N>=1$ threshold is used to subtract the substructure. After point
source subtraction and substructure excision, hydrostatic equilibrium
should usually be valid in the remaining sector (more details see
Sect.~\ref{s:discu}).

\subsection{Spectral analysis}
\label{s:bkgspe}

We use the XSPEC v11.3.1 software for the spectral analysis. A double
background subtraction procedure can be applied in two ways in the
spectral analysis. One approach was firstly described in Zhang et
al. (2004a, 2005a, hereafter DBS I). The other method was well
illustrated in Pratt \& Arnaud (2002, hereafter DBS II). We applied
both methods to perform the data reduction and obtained similar
results using these two approaches.

For a given region of interest covering cluster emission, the spectrum
is extracted from the background observations in the same detector
coordinates as for the target. Both the response matrix file (rmf) and
auxiliary response file (arf) are used to recover the correct spectral
shape and normalization of the cluster emission component. The
following is usually taken into account for the rmf and arf, (i) a
pure redistribution matrix giving the probability that a photon of
energy E, once detected, will be measured in data channel PI, (ii) the
quantum efficiency (without any filter, which, in XMM-Newton
calibration, is called closed filter position) of the CCD detector,
(iii) filter transmission, (iv) geometric factors such as bad pixel
corrections and gap corrections (e.g. around 4\% for MOS), and (v)
telescope effective area as a function of photon energy. The
vignetting correction to effective area for off-axis observations can
be accounted for either in the arf or in the event lists by a weight
column created by ``evigweight''.

In DBS I, the rmf corresponds to (i) and (ii), and the arf corresponds
to (iii), (iv), (v) as well as the vignetting correction. A source
spectrum is extracted from the outer region of the target
observations. A background spectrum is extracted from the background
observations in the same detector coordinates.  Using only the rmf,
the residual background spectrum is found after subtracting the
background spectrum from the source spectrum in the outer region.  We
assume as an extreme case that the residual background mainly consists
of the BNVIGs, e.g. soft protons. The vignetting effect can thus be
ignored for the residual spectrum. It is modeled by a ``powerlaw/b''
model (``powerlaw/b'' ,a power law background model which is convolved
with the instrumental redistribution matrix but not with the effective
area) in XSPEC limited to 0.4--15~keV. This model is used to account
for the residual background in the spectral analysis over the whole
energy range by applying a combined model of
``wabs$*$mekal$+$powerlaw/b'' (``mekal'', an emission model for hot
diffuse gas, cf. Mewe et al. 1985, 1986; Arnaud \& Rothenflug 1985;
Arnaud \& Raymond 1992; Kaastra 1992; Liedahl et al. 1995; ``wabs'', a
model considering the Galactic absorption) in XSPEC in the fitting
procedure. The correct shape of the background component is recovered
in the fit when the normalization of the ``powerlaw/b'' model is
scaled to the area of the given region. The uncertainties of the
fitting parameters of the ``powerlaw/b'' model are not introduced in
the spectral analysis. This might result in an underestimate of the
temperature uncertainty in the spectral analysis using DBS I. However,
the advantage of DBS I is that the shape of the residual spectrum is
conserved during the procedure.

\begin{figure*}
\begin{center}
\includegraphics[width=9cm,angle=0]{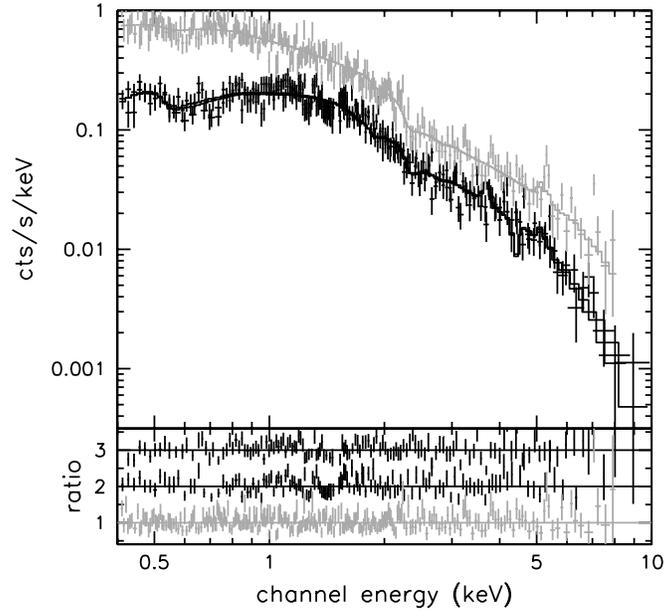}
\end{center}
\caption{XMM-Newton spectra (pn in grey and MOS in black) of
RXCJ0014.3$-$3022 of the 0.1--0.5~$r_{500}$ region fitted by the
combined ``wabs$*$mekal'' model. The ratios of the observational data
to the models are in the lower parts of the panels (offset zero for
pn, $+1$ for MOS1, $+2$ for MOS2). \label{f:spe} }
\end{figure*}

In DBS II, all spectra are extracted considering the vignetting
correction by a weighted column in the event list produced by
``evigweight''. The on-axis rmf and arf are co-created to account
for (i) to (v).  The target spectrum is extracted from the region
of interest. The first-order background spectrum is given by a
spectrum from the background observations which is extracted in
the same detector coordinates as for the source spectrum, and
which is scaled using the ctr ratio of the target and background
limited to 10--12~keV (12--14~keV) for MOS (pn). The second-order
background spectrum, the residual background spectrum, is prepared
as follows. A source spectrum is extracted from the outer region
of the target observations and its background spectrum from the
same outer region of the background observations. The background
spectrum is scaled using the ctr ratio of the target and
background limited to 10--12~keV (12--14~keV) for MOS (pn), and is
subtracted from the source spectrum to obtain a residual
background spectrum. This residual background spectrum is
normalized to the area of the region of interest as a second-order
background spectrum. Both the first-order and second-order
background spectra are subtracted from the target spectrum. A
combined model of ``wabs$*$mekal'' is then used with the on-axis
arf and rmf in XSPEC for the fitting (Fig.~\ref{f:spe}).

Basically, the vignetting correction is applied on the residual
background in DBS II and not applied in DBS I. The former method is
thus in principle better when the residual is dominated by the BVIGs
(e.g. the CXB residual component), while the latter is better when the
residual is dominated by the BNVIGs. The two approaches provide
consistent results at the 1-$\sigma$ confidence level which indirectly
tested the two approaches. A small discrepancy was found mostly in the
last bin but showed no trend of being higher for one particular
approach. There is no preferred indication of one of the two
approaches. Additionally, the DBS~II approach is simple in, (i)
modeling the residual background, by applying the residual spectrum,
instead of looking for an acceptable ($\chi^2 < 2$) power law model
fit; (ii) allowing negative residual components, and (iii) generating
one on-axis arf working for all the spectra extracted from different
annuli. Therefore, we adopt the measurements using DBS II to
illustrate the further computations.

\subsection{Image analysis}

The 0.5--2~keV band is used to derive the surface brightness
profiles (also see Zhang et al. 2005a). This ensures an almost
temperature-independent X-ray emission coefficient over the
expected temperature range. The vignetting correction to effective
area is accounted for using a weight column in the event lists
created by ``evigweight''. Geometric factors such as bad pixel
corrections are accounted for in the exposure maps. The width of
the radial bins is $2^{\prime \prime}$. An azimuthally averaged
surface brightness profile of CDFS is derived in the same detector
coordinates as for the targets. The ctr ratios of the targets and
CDFS in the 10--12~keV band and 12--14~keV band for MOS and pn,
respectively, are used to scale the CDFS surface brightness. The
residual background in the 0.5--2~keV band is introduced by using
the determined residual spectrum in the spectral analysis. The
background-subtracted and vignetting-corrected surface profiles
for three detectors are added into a single profile, and re-binned
to reach a significant level of at least 3-$\sigma$ in each
annulus. We take into account a 10\% uncertainty of the scaled
CDFS background and residual background.

\subsection{PSF and de-projection}
\label{s:deproj}

Using the XMM-Newton PSF calibrations by Ghizzardi (2001) we estimated
the redistribution fraction of the flux. We found 20\% for bins with
width about $0.5^{\prime}$ and less than 10\% for bins with width
greater than $1^{\prime}$ neglecting energy dependent effects. The PSF
blurring can not be completely considered for the spectral analysis as
it is done for the image analysis because of the limited photon
statistic. For such distant clusters, the PSF effect is only
important within $0.3r_{\rm 500}$ and introduces an added uncertainty
to the final results of the temperature profiles. This has to be
investigated using deeper exposures with better photon statistic.

The projected temperature is the observed temperature from a
particular annulus, containing in projection the foreground and
background temperature measurements. Under the assumption of spherical
symmetry, the gas temperature in each spherical shell is derived by
de-projecting the projected spectra. In this procedure, the inner
shells contribute nothing to the outer annuli. The projected spectrum
in the outermost annulus is thus equal to the spectrum in the
outermost shell. The projected spectrum in the neighboring inner
annulus has contributions from all the spectra in the shell at the
radius of this annulus and in the outer shells as shown in Suto et
al. (1998). In XSPEC, a combined fit to the projected spectra measured
in all annuli simultaneously gives the contribution of each spherical
shell to each annulus and provides the de-projected measurements of
the temperature and metallicity.

In the imaging analysis, we correct the PSF effect by fitting the
observational surface brightness profile with a surface brightness model
convolved with the empirical PSF matrices (Ghizzardi 2001). The
surface brightness model is calculated by the projection of the radial
electron density profile.

\section{X-ray properties}
\label{s:result}

The primary parameters of the REFLEX-DXL clusters are given in
Table~\ref{t:global}.

\subsection{Density contrast}

The mean cluster density contrast, $\Delta$, is the average density with
respect to the critical density, $\rho_{\rm c}(z)=\rho_{\rm c0}
E^2(z)$, where $E^2(z)=\Omega_{\rm
m}(1+z)^3+\Omega_{\Lambda}+(1-\Omega_{\rm
m}-\Omega_{\Lambda})(1+z)^2$. $r_{\Delta}$ is the radius within which
the density contrast is $\Delta$. $M_{\Delta}$ is the total mass within
$r_{\Delta}$. For $\Delta=500$, $r_{500}$ is the radius within which
the density contrast is 500 and $M_{500}$ is the total mass within
$r_{500}$. 

\subsection{Metallicity and temperature}
\label{s:kt}

We found that the X-ray determined redshifts agree with the optically
measured redshifts (also in Zhang et al. 2004a). We therefore fixed
the redshift to the optical redshift in the further analysis. The
spectrum is fitted by a combined ``wabs$*$mekal'' model with fixed
Galactic absorption (Dickey \& Lockman 1990) and redshift (B\"ohringer
et al. 2004a).

\begin{figure*}
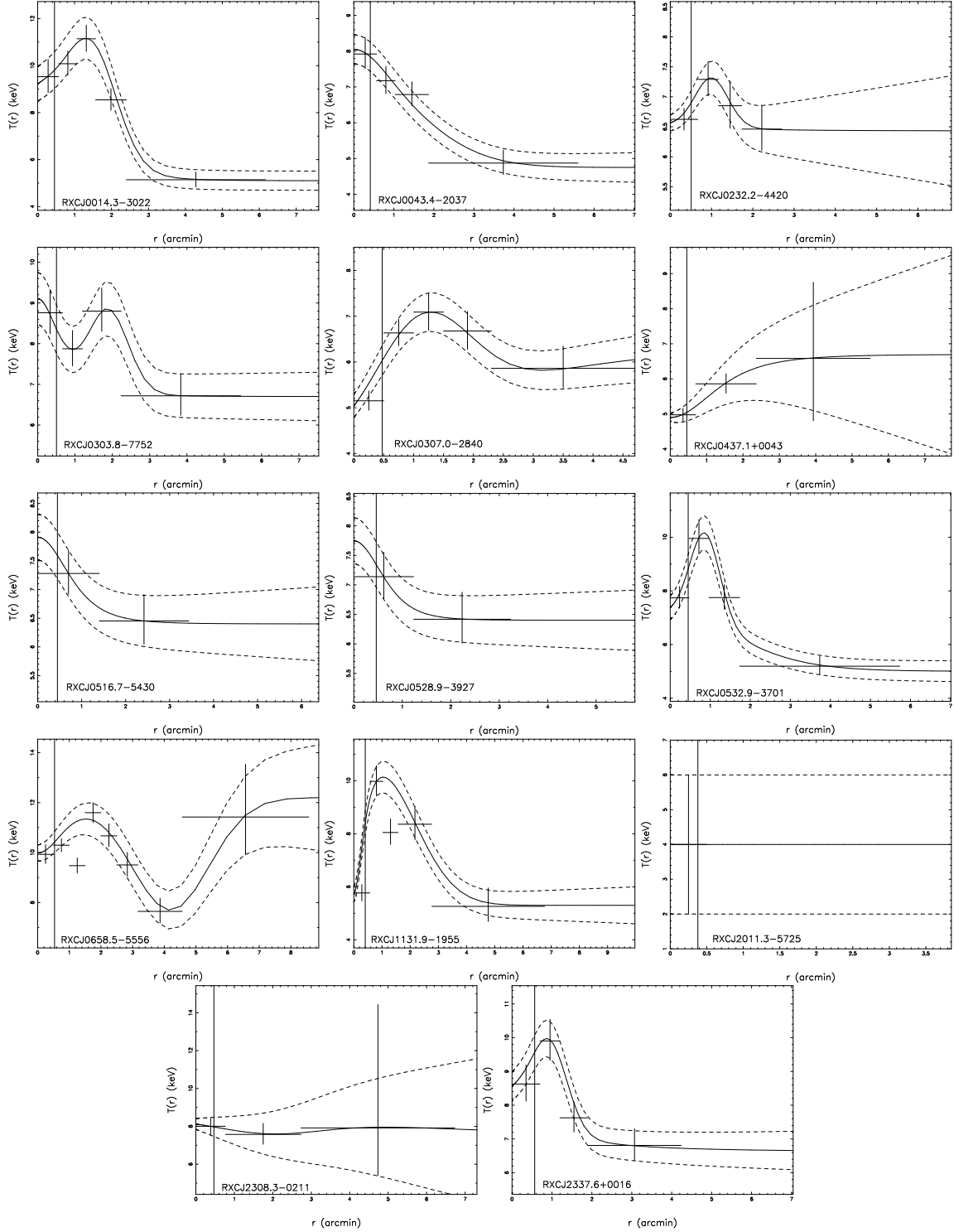

\begin{center}
\includegraphics[angle=270,width=5cm]{3650f2a.ps}
\includegraphics[angle=270,width=5cm]{3650f2b.ps}
\includegraphics[angle=270,width=5cm]{3650f2c.ps}

\includegraphics[angle=270,width=5cm]{3650f2d.ps}
\includegraphics[angle=270,width=5cm]{3650f2e.ps}
\includegraphics[angle=270,width=5cm]{3650f2f.ps}

\includegraphics[angle=270,width=5cm]{3650f2g.ps}
\includegraphics[angle=270,width=5cm]{3650f2h.ps}
\includegraphics[angle=270,width=5cm]{3650f2i.ps}

\includegraphics[angle=270,width=5cm]{3650f2j.ps}
\includegraphics[angle=270,width=5cm]{3650f2k.ps}
\includegraphics[angle=270,width=5cm]{3650f2l.ps}

\includegraphics[angle=270,width=5cm]{3650f2m.ps}
\includegraphics[angle=270,width=5cm]{3650f2n.ps}

\end{center}
\caption{De-projected temperature profiles of the REFLEX-DXL
clusters. The temperature profiles are approximated by the
parameterization $T(r)=T_3
\exp[-(r-T_1)^2/T_2]+T_6(1+r^2/T_4^2)^{-T_5}+T_7$ crossing all the
data points (solid). The 1-$\sigma$ confidence intervals are
approximated by $T(r)[1 \pm (T_8+T_9 r)]$ (dashed). The vertical line
denotes $0.1 r_{500}$. \label{f:ktcom}}
\end{figure*}

Temperature profiles can provide a useful means to study the
thermodynamical history of galaxy clusters. XMM-Newton (and
Chandra), in contrast to earlier telescopes, provides a less
energy-dependent, and smaller, PSF. It is thus more reliable to
study cluster temperature profiles with XMM-Newton. A systematic
spectral analysis was performed in annuli. We re-binned the data
to contain around 500--550 net counts per annulus for MOS1 in the
2--7~keV band. In XSPEC, a combined fit of ``wabs$*$mekal$*$projct''to
the projected spectra measured in all annuli simultaneously gives
the contribution of each spherical shell to each annulus and
provides the de-projected measurements of the temperature and also
metallicity. In Fig.~\ref{f:ktcom} we show the de-projected
temperature profiles of the REFLEX-DXL clusters. The temperature
profiles are approximated by the parameterization $T(r)=T_3
\exp[-(r-T_1)^2/T_2]+T_6(1+r^2/T_4^2)^{-T_5}+T_7$ crossing all the
data points. The temperature measure uncertainties are
approximated by $T(r)\times (T_8+T_9r)$ and are propagated for
individual data points by Monte Carlo simulations in the further
analysis.

Vikhlinin et al. (2005) used the weighting of the 3-D temperature
profile and averaged the temperature profile in a certain radial
range. For the REFLEX-DXL sample, the radial temperature profiles are
almost self-similar above 0.1--0.2$r_{500}$. The REFLEX-DXL data set
shows significantly low $S/N$ above $0.5 r_{500}$.  Therefore, the
volume averaged radial temperature profile of 0.1--0.5~$r_{500}$ is
chosen as the global temperature $T$ listed in
Table~\ref{t:global}. Similarly, the metallicity profile of
0.1--0.5~$r_{500}$ is used to obtain the global metallicity $Z$ also
listed in Table~\ref{t:global}. We found an average of $0.24 \pm
0.03$~$Z_{\odot}$ for the global metallicities of the REFLEX-DXL
sample excluding RXCJ2011.3$-$5725. This agrees with the value of
0.21~$Z_{\odot}$ in Allen \& Fabian (1998). It is also consistent with
the averaged metallicity $\overline{Z}=0.21^{+0.10}_{-0.05} Z_{\odot}$
for 18 distant ($0.3<z<1.3$) clusters in Tozzi et al. (2003).

Based on the flared observations of RXCJ2011.3$-$5725, we measured a
global spectral temperature ($\sim 4 \pm 2$~keV) with a fixed
metallicity of 0.3~$Z_{\odot}$ within $0.5^{\prime}$ radius.

\subsection{Surface brightness}
\label{s:sx}

\begin{figure*}
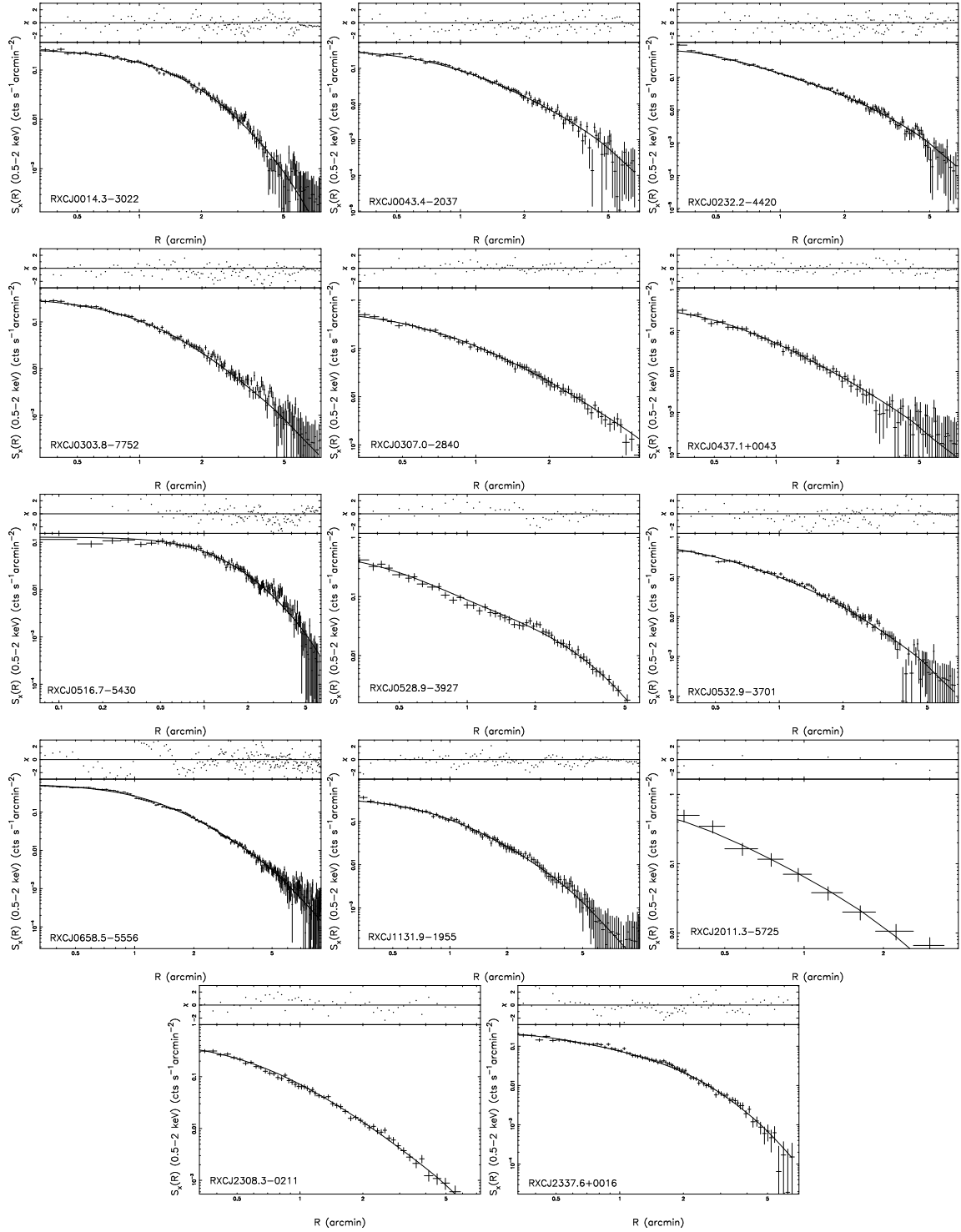

\begin{center}
\includegraphics[angle=270,width=5cm]{3650f3a.ps}
\includegraphics[angle=270,width=5cm]{3650f3b.ps}
\includegraphics[angle=270,width=5cm]{3650f3c.ps}

\includegraphics[angle=270,width=5cm]{3650f3d.ps}
\includegraphics[angle=270,width=5cm]{3650f3e.ps}
\includegraphics[angle=270,width=5cm]{3650f3f.ps}

\includegraphics[angle=270,width=5cm]{3650f3g.ps}
\includegraphics[angle=270,width=5cm]{3650f3h.ps}
\includegraphics[angle=270,width=5cm]{3650f3i.ps}

\includegraphics[angle=270,width=5cm]{3650f3j.ps}
\includegraphics[angle=270,width=5cm]{3650f3k.ps}
\includegraphics[angle=270,width=5cm]{3650f3l.ps}

\includegraphics[angle=270,width=5cm]{3650f3m.ps}
\includegraphics[angle=270,width=5cm]{3650f3n.ps}

\end{center}
\caption{Surface brightness profiles of the REFLEX-DXL clusters and
their best fits. Residuals scaled by the data uncertainties are
plotted in the upper part of the diagrams.
\label{f:sx0}}
\end{figure*}

A $\beta$-model (e.g. Cavaliere \& Fusco-Femiano 1976; Jones \& Forman
1984) is often used to describe electron density profiles. To obtain
an acceptable fit for all clusters in this sample, we adopt a
double-$\beta$ model, $n_{\rm e}(r)=n_{\rm e01}(1+r^2/r_{\rm
  c1}^2)^{-3\beta/2}+n_{\rm e02}(1+r^2/r_{\rm c2}^2)^{-3\beta/2}$. The
X-ray surface brightness profile model is linked to the radial
profile of the ICM electron number density as an integral
performed along the line-of-sight,
\begin{equation}
S_{\rm X}(R)\propto\int_{R}^{\infty} n_{\rm e}^2 d\ell.
\end{equation}
We fit the observed surface brightness profile by this integral
convolved with the PSF matrices (Fig.~\ref{f:sx0}) and obtain the
parameters of the double-$\beta$ model of the electron density
profile. The fit was performed within the truncation radius ($r_{\rm
t}$, see Table~\ref{t:global}) corresponding to a $S/N$ of 3 of the
observational surface brightness profile. The truncation radii,
$r_{\rm t}$, are about or above $r_{500}$ for the REFLEX-DXL
clusters. The X-ray bolometric luminosity (here we use the
0.01--100~keV band) is given by $L_{\rm X}^{\rm bol} \propto \int
\widetilde{\Lambda}(r) n^2_{\rm e}(r) dV$, practically an integral of
the X-ray surface brightness to infinity (In practice, $20^{\prime}$
is used for the integral upper radial limit). We applied a power law
to fit the surface brightness in the outskirts ($>0.4r_{500}$). More
details about the slope of the the outskirts can be found in
Sect.~\ref{s:infall}.

\subsection{Cooling time}

The cooling time is derived by the total energy of the gas divided by
the energy loss rate
\begin{equation}
t_{\rm c} = \frac{2.5 n_{\rm g}T}{n_{\rm e}^2\widetilde{\Lambda}}
\label{eq:tcool}
\end{equation}
where $\widetilde{\Lambda}$, $n_{\rm g}$, $n_{\rm e}$ and $T$ are the
radiative cooling function, gas number density, electron number
density and temperature, respectively. We compute the upper limit of
the age of the cluster as an integral from the cluster redshift $z$ up
to $z=100$. Cooling regions are those showing cooling time less than
the upper limit of the cluster age. The boundary radius of such a
region is called the cooling radius. The cooling radius is zero when
the cooling time is larger than the cluster age. The cooling times and
cooling radii are given in Table~\ref{t:catalog1}.

\subsection{Gas entropy}

The entropy is the key to an understanding of the thermodynamical
history since it results from shock heating of the gas during cluster
formation. The observed entropy is generally defined as $S=T n_{\rm
e}^{-2/3}$ for galaxy cluster studies (e.g. Ponman et al. 1999), and
it scales with the cluster temperature. An excess above the scaling
law indicates non-gravitational heating effects by sources such as the
feedback from super-novae (SN) and AGNs (e.g. Lloyd-Davies et
al. 2000). Radiative cooling can also raise the entropy of the ICM
(e.g Pearce et al. 2000) or produce a deficit below the scaling law.
In this sample, the clusters appearing more relaxed are identical to
those showing lower central entropy values. For the REFLEX-DXL
clusters, the most inner temperature data points are measured at $\sim
0.1r_{500}$. This leads to the fact that the entropies at $0.1r_{500}$
($S_{\rm 0.1 r_{500}}$) are resolved and can be used as the values of
the central entropies as shown in the entropy radial distributions
(Fig.~\ref{f:en}). At $0.1 r_{200}$ (about $0.18 r_{500}$), the
entropy increases as a function of radius and is thus significantly
larger than the central entropy.

\subsection{Mass distribution}
\label{s:massdis}

We assume that the intra-cluster gas is in hydrostatic equilibrium
within the gravitational potential dominated by DM. The ICM can
thus be used to trace the cluster potential. Under the assumption
of spherical symmetry, the cluster mass is calculated from the
X-ray measured ICM density and temperature distributions,
\begin{equation}
\frac{1}{\mu m_{\rm p} n_{\rm e}(r)}\frac{d[n_{\rm e}(r) T(r)]}{dr}=
  -\frac{GM(<r)}{r^2}~,
\label{e:hyd}
\end{equation}
where $\mu=0.62$ is the mean molecular weight per hydrogen
atom. Following the Monte Carlo simulation method (e.g. Neumann \&
B\"ohringer 1995), we use a set of input parameters of the
approximation functions, in which $\beta$, $n_{\rm e0i}$, $r_{\rm ci}$
($i=1,2$) are for the gas density radial profile $n_{\rm e}(r)$ and
$T_{\rm i}$ ($i=1,...,7$) are for the temperature radial profile
$T(r)$, respectively, to compute the cluster mass. The individual data
point uncertainties are propagated by Monte Carlo simulations.
We used the measured mass profile to estimate $M_{500}$ and
$r_{500}$.

\begin{figure*}
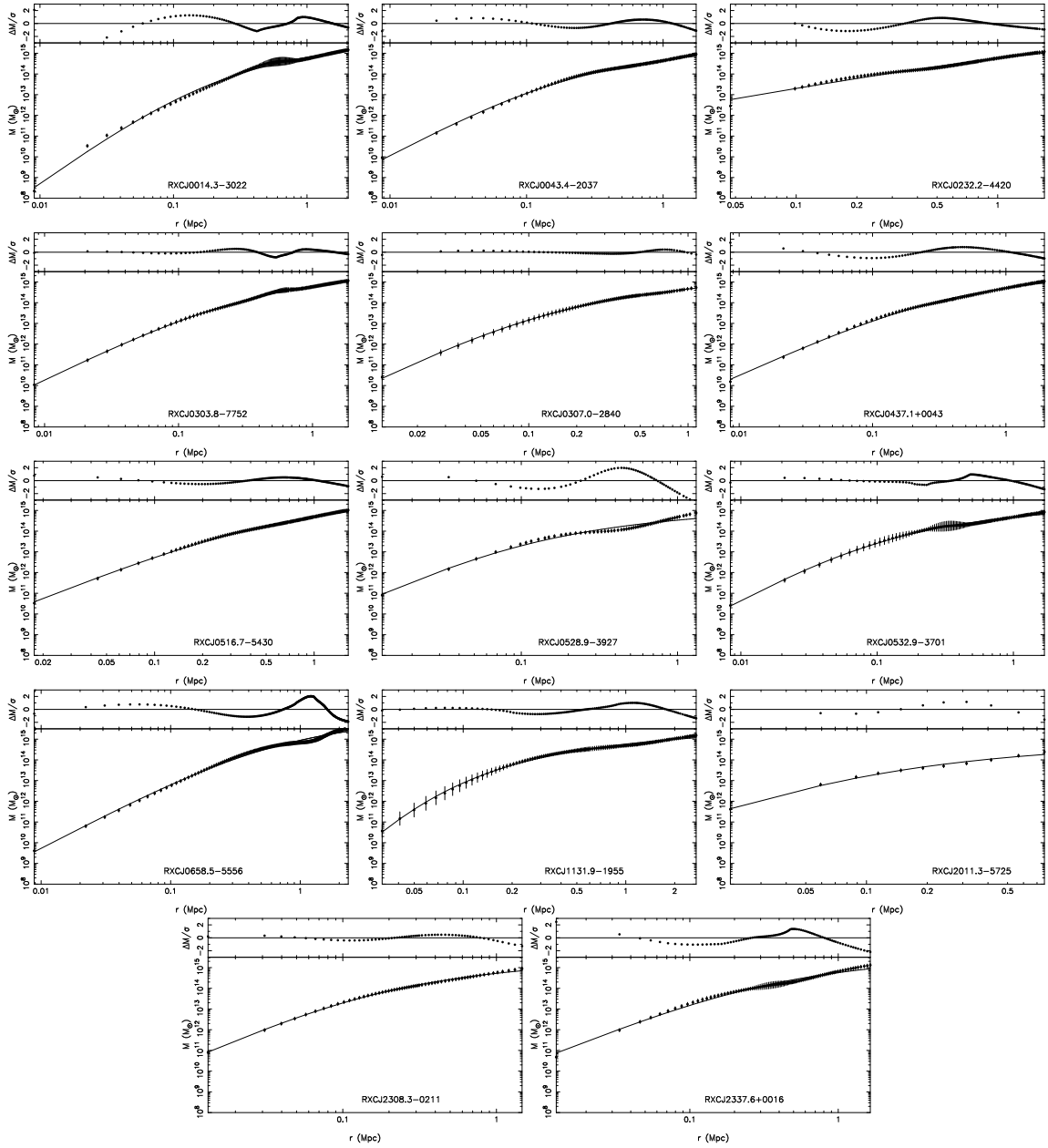

\begin{center}
\includegraphics[angle=270,width=5cm]{3650f5a.ps}
\includegraphics[angle=270,width=5cm]{3650f5b.ps}
\includegraphics[angle=270,width=5cm]{3650f5c.ps}

\includegraphics[angle=270,width=5cm]{3650f5d.ps}
\includegraphics[angle=270,width=5cm]{3650f5e.ps}
\includegraphics[angle=270,width=5cm]{3650f5f.ps}

\includegraphics[angle=270,width=5cm]{3650f5g.ps}
\includegraphics[angle=270,width=5cm]{3650f5h.ps}
\includegraphics[angle=270,width=5cm]{3650f5i.ps}

\includegraphics[angle=270,width=5cm]{3650f5j.ps}
\includegraphics[angle=270,width=5cm]{3650f5k.ps}
\includegraphics[angle=270,width=5cm]{3650f5l.ps}

\includegraphics[angle=270,width=5cm]{3650f5m.ps}
\includegraphics[angle=270,width=5cm]{3650f5n.ps}

\end{center}
\caption{Measured mass profiles of the REFLEX-DXL clusters and
their best fits by the extended-NFW models. Residuals scaled by the
data uncertainties are plotted in the upper part of the diagrams.
\label{f:exnfw1}}
\end{figure*}

The NFW model (e.g. Navarro et al. 1997, 2004, NFW) cannot provide an
acceptable fit for the observed mass profiles. We therefore adopt the
best fit of an extended-NFW model (e.g. Hernquist 1990; Zhao 1996;
Moore et al. 1999), $\rho_{\rm DM}(r)=\rho_{\rm s} (r/r_{\rm
s})^{-\alpha}(1+r/r_{\rm s})^{\alpha-3}$, where $\rho_{\rm s}$ and
$r_{\rm s}$ are the the characteristic density and scale of the halo,
respectively. In Fig.~\ref{f:exnfw1}, we show the
measured mass profiles and their best extended-NFW model fits. We
derive the concentration parameters ($c_{500}$) of the DM
distributions of the REFLEX-DXL clusters (see Table~\ref{t:catalog2})
using the extended-NFW model. The difference in the concentration
parameters using different models can be large. For instance, an
extended-NFW profile fitted to the numerical simulations of Moore et
al (1999) gives a concentration parameter 50\% higher than the
concentration parameter given by the NFW model. In general, the
theoretical and observed concentration--mass relations are compared
when the same models (e.g. NFW) are considered. However, the
REFLEX-DXL data cannot be well fitted by the NFW model. We thus
cannot directly compare our results to the published observations
(e.g. Pointecouteau et al 2005) and simulations (e.g. Dolag et
al. 2004) which are based on the NFW model. The best fit of the REFLEX-DXL
sample gives the slope of $1.5\pm0.2$ and the normalization of
$10^{2.1\pm0.2}$ (Fig.~\ref{f:cmass}). If we fixed the slope parameter to
$-0.102$ for the concentration--mass relation as found in Dolag et
al. (2004), the best fit of the REFLEX-DXL sample gives the
normalization of $8.7\pm7.8$.

\subsection{Gas mass fraction distribution}

The gas mass fraction is an important parameter for cluster physics,
e.g. heating and cooling processes, and cosmological applications
using galaxy clusters (e.g. Vikhlinin et al. 2002; Allen et
al. 2004). The gas mass fraction distribution is defined to be $f_{\rm
gas}(<r)=M_{\rm gas}(<r)/M(<r)$.

As shown in Fig.~\ref{f:fg}, the gas mass fractions increase as a
function of radius in the $r_{3000}<r<r_{1000}$ region. This indicates
that the DM distribution is more concentrated in the center than the
gas distribution. We derived an average gas mass fraction of $0.116
\pm 0.007$ at $r_{\rm 500}$. This agrees with the gas mass fractions
found for many massive clusters showing temperatures greater than
5~keV (e.g. Mohr et al. 1999). The gas mass fractions around
$r_{2500}$ show the smallest scatter, $0.100 \pm 0.007$, and the
values are similar to the measurements of Allen et al. (2002) based on
Chandra observations of 7 clusters yielding $f_{\rm gas}\sim
0.105$--$0.138h^{-3/2}_{70}$. At $r_{200}$, the extrapolated gas mass
fractions show consistency with the measurements of Sanderson et
al. (2003) based on ASCA/GIS, ASCA/SIS and ROSAT/PSPC observations of
66 clusters yielding $f_{\rm gas}=0.13\pm 0.01 h^{-3/2}_{70}$, the
measurements of Ettori et al. (2002a) based on BeppoSAX observations
of 22 nearby clusters, and the gas mass fraction for A1413 (Pratt \&
Arnaud 2002) at $z=0.143$ based on XMM-Newton observations yielding
$f_{\rm gas}\sim 0.12h^{-3/2}_{70}$. As expected, the gas mass
fraction distributions of the REFLEX-DXL clusters are lower than the
universal baryon fraction, $f_{\rm b}=\Omega_{\rm b}/\Omega_{\rm
m}=0.167 \pm 0.014$, based on the recent WMAP measurements,
$h=0.71^{+0.04}_{-0.03}$, $\Omega_{\rm b}~ h^2=0.022 \pm 0.001$ and
$\Omega_{\rm m}~ h^2=0.132^{+0.008}_{-0.009}$ (e.g. Spergel et
al. 2003; Hansen et al. 2004). This is because the baryons in galaxy
clusters reside mostly in hot gas together with a fraction
(15\%~$f_{\rm gas}$) in stars as implied from simulations (e.g. Eke et
al. 1998; Kravtsov et al. 2005). In principle, $\Omega_{\rm m}$ can be
determined from the baryon fraction, $f_{\rm b}=f_{\rm gas}+f_{\rm
gal}$, in which a contribution from stars in galaxies is given by
$f_{\rm gal}=0.02\pm 0.01 h_{50}^{-1}$ (White et al. 1993). The gas
mass fractions, $\sim 0.116\pm 0.007$, at $r_{500}$ of the REFLEX-DXL
clusters support a low matter density Universe as also shown in recent
studies (e.g. Allen et al. 2002; Ettori et al. 2003; Vikhlinin et
al. 2003).

\section{Scaling relations}
\label{s:scale}

Simulations (e.g. Navarro et al. 1997, 2004) suggest a
self-similar structure for galaxy clusters in hierarchical
structure formation scenarios. The scaled profiles of the X-ray
properties and their scatter can be used to quantify the
structural variations. This is a probe to test the regularity of
galaxy clusters and to understand their formation and evolution.
The accuracy of the determination of the scaling relations,
limited by how precise the cluster mass and other global
observable can be estimated, is of prime importance for the
cosmological applications of clusters of galaxies.

Because the observational truncation radii ($r_{\rm t}$) in the
surface brightness profiles are about or slightly above $r_{500}$, we
use $r_{500}$ for radial scaling.

The following redshift evolution corrections (e.g. Ettori et al. 2004)
are usually used to check the dependence on the evolution of the
cosmological parameters,

$S_{\rm X} \cdot E^{-3}(z) \; (\Delta_{c,z}/\Delta_{c,0})^{-1.5}
\propto f(T)$,

$S \cdot E^{4/3}(z) \; (\Delta_{c,z}/\Delta_{c,0})^{2/3} \propto f(T)$,

$L \cdot E^{-1}(z) \; (\Delta_{c,z}/\Delta_{c,0})^{-0.5} \propto f(T)$,

$M \cdot E(z) \; (\Delta_{c,z}/\Delta_{c,0})^{0.5} \propto f(T)$,

$M_{\rm gas} \cdot E(z) \; (\Delta_{c,z}/\Delta_{c,0})^{0.5} \propto f(T)$,

where $\Delta_{\rm c,z}=18\pi^2+82[\Omega_{\rm m,z}-1]-39[\Omega_{\rm
m,z}-1]^2$ for a flat Universe and $\Omega_{\rm m,z}$ is the cosmic
density parameter at redshift $z$.

\subsection{Scaled temperature profiles}

Studies of the cluster temperature distributions (e.g. Markevitch et
al. 1998; De Grandi \& Molendi 2002) indicate a steep decline beyond
an isothermal center. We scaled the radial temperature profiles by the
global temperature $T$ and $r_{500}$ as shown in
Fig.~\ref{f:scalet}. An average temperature profile is derived by
averaging the 1-$\sigma$ boundary of the scaled radial temperature
profiles of the REFLEX-DXL clusters. As shown in Fig.~\ref{f:scalet},
we found a closely self-similar behaviour. Up to $0.3r_{\rm 500}$, we
observed an almost constant temperature distribution with a
temperature peak at around 0.2~$r_{\rm 500}$. Four clusters
(RXCJ0232.2$-$4420, RXCJ0307.0$-$2840, RXCJ0437.1$+$0043 and
RXCJ0528.9$-$3927) show relatively cool cluster cores. A temperature
profile decreasing down to 80\% of the peak value has been observed
outside $0.3 r_{\rm 500}$ for the average of the REFLEX-DXL
clusters. Three clusters (RXCJ0437.1$+$0043, RXCJ0658.5$+$5556, and
RXCJ2308.3$+$0211) show temperatures rising with relatively large
error bars in their temperature profiles. The average temperature
profile can decline down to 50\% of the peak value when these three
clusters are excluded. This average temperature profile is consistent
with the average profiles from ASCA in Markevitch et al. (1998),
BeppoSAX in De Grandi \& Molendi (2002) and Chandra (using an assumed
uncertainty of 20\% of the averaged temperature profile as an
approximate illustration) in Vikhlinin et al. (2005) within the
observational dispersion. A similarly universal temperature profile is
indicated by simulations (e.g. Borgani et al. 2004; Borgani 2004).

\subsection{Scaled surface brightness profiles}

We scaled the surface brightness profiles using the self-similar
scaling, $S_{\rm X} \propto T^{0.5}$, and the empirical scaling,
$S_{\rm X} \propto T^{1.38}$, respectively, as described in
e.g. Arnaud et al. (2002). Both scaled profiles show a less scattered
self-similar behaviour at $r>0.2 r_{500}$ (see
Fig.~\ref{f:scalesx}). The core radii populate a broad range of
values, 0.03--0.2~$r_{500}$.

\subsection{Scaled entropy profiles}
\label{s:en}

According to the standard self-similar model the entropy scales as $S
\propto T$ (e.g. Pratt \& Arnaud 2005).  Ponman et al. (2003)
suggested to scale the entropy as $S \propto T^{0.65}$. We investigate
the entropy--temperature relation ($S$--$T$) using $S_{\rm 0.1
r_{500}}$ to represent the central entropies. Four REFLEX-DXL clusters
(RXCJ0232.2$-$4420, RXCJ0307.0$-$2840, RXCJ0437.1$+$0043 and
RXCJ0528.9$-$3927) show significantly lower central entropies compared
to the $S$--$T$ scaling law. These clusters have relatively cool
cluster cores as observed in their temperature profiles. Neglecting
the resolution problem, we found that the radiative cooling effect
introduces a significant scatter in the $S$--$T$ scaling relation in
terms of a lower ($\sim 7\%$) normalization. Furthermore, 3
pronounced merger clusters (RXCJ0014.3$-$3022, RXCJ0516.7$-$5430 and
RXCJ2337.6$+$0016) show relatively higher central entropies. This
indicates that mergers may also introduce some scatter in the $S$--$T$
relation but in terms of a slightly higher ($\sim 10\%$)
normalization. The central entropy can thus be used not only as a
mechanical educt of the non-gravitational process, but also as an
indicator of the merger stage. Excluding the above 7 clusters
discussed, we performed a best fit for the $S$--$T$ relation of the
remaining 6 clusters giving $S\propto T^{0.63\pm0.15}$. This fit
agrees with the $S$--$T$ relation of the Birmingham-CfA clusters
(Ponman et al. 2003). Therefore, we scaled the radial entropy profile
using the empirically determined scaling (Ponman et al. 2003),
$S\propto T^{0.65}$, and $r_{500}$. As shown in Fig.~\ref{f:en}, the
scaled entropy profiles of the REFLEX-DXL clusters agree with the
scaled entropy profiles of the Birmingham-CfA clusters in Ponman et
al. (2003) and the clusters in Pratt et al. (2006) in the same
temperature range (6--20~keV) within the observational
dispersions. The least scatter ($\sim 30\%$) of the entropy profiles
is around 0.2--0.3$r_{500}$. The combined entropy profiles give the
best fit, $S(r)\propto r^{0.99\pm 0.06}$, above $0.1r_{\rm 500}$.  A
similar power law as $S\propto r^{0.97}$ was found by Ettori et
al. (2002b) and $S\propto r^{0.95}$ by Piffaretti et al. (2005). The
slope of the entropy profiles is shallower at the 1-$\sigma$
confidence level than the predicted slope from a spherical accretion
shock model, $S\propto r^{1.1}$ (e.g. Kay 2004).
 
\subsection{Scaled total mass and gas mass profiles}

The mass profiles were scaled with respect to $M_{\rm 500}$ and
$r_{\rm 500}$, respectively (Fig.~\ref{f:scaleym}). We found the least
scatter at radii above 0.2--0.3~$ r_{\rm 500}$. In the inner parts
($<0.2 r_{\rm 500}$), the mass profiles vary significantly with the
cluster central dynamics (Fig.~\ref{f:scaleym}). For the clusters
showing cooling cores, the mass distributions are relatively cuspidal,
while for the merger clusters, less concentrated mass distributions
are observed. The scaled gas mass profiles appear more self-similar
than the scaled mass profiles, especially at radii above
0.2--0.3~$r_{\rm 500}$.

\subsection{ROSAT and XMM-Newton luminosities}

Substructure is often observed in galaxy clusters and the frequency of
its occurrence has for example been estimated to be of the order of
about $52\pm 7$\% (Schuecker et al. 2001). The high resolution
XMM-Newton data allow us to identify the substructures and point-like
sources better than what was possible with the earlier X-ray
telescopes. Subtracting the substructure contribution also from the
ROSAT measured cluster luminosity, we found a good agreement between
the ROSAT and XMM-Newton luminosity for the REFLEX-DXL clusters
(Fig.~\ref{f:lxmmrosat}). The XMM-Newton data provide a more reliable
and complete detection of substructures and point-like
sources. Therefore we make use of this capability to obtain a better
approximation to spherical symmetry and dynamical equilibrium of the
main, largely relaxed component by excluding the substructures and
point-like sources.  The properties of the main cluster component are
more representative for investigating various scaling relations of
regular galaxy clusters, such as the $L$--$T$ relation.

\subsection{Scaling relations}
\label{s:lmt}

To use the temperature/mass function of this unbiased,
flux-limited and almost volume-complete sample to constrain
cosmological parameters, it is important to calibrate the scaling
relations between the X-ray luminosity, temperature and
gravitational mass. The scaling relations can generally be
parameterized by a power law.

The scatter describes the dispersion between the observational data
points and the best fit. The scatter in the scaling relation is
strongly dependent on the temperature measurement uncertainty. The
massive clusters in a narrow temperature range provide an important
means to constrain the normalization of the scaling relations such as
the $M$--$T$ relation. We collected recently published scaling
relations and compared them to the results of this work in
Table~\ref{t:mtx_lite} and Figs.~\ref{f:mt}--\ref{f:fgratio}. The
scatter in the scaling relations can partially be explained by
variation of the cluster morphology. Comparing the REFLEX-DXL sample
to the nearby samples, we found that the evolution of the scaling
relations are accounted for by the redshift evolution given in
Sect.~\ref{s:scale}. We obtained an overall agreement with the recent
studies of the scaling relations within the observational
dispersion. This fits into the general opinion that galaxy clusters
are self-similar up to $z\sim 1$ (e.g. Arnaud 2005).

To determine the normalization, the slope of the $M_{500}$--$T$
relation is fixed to 1.5 in the fitting procedure as also used in many
published results (e.g. Evrard et al. 1996, Ettori et al. 2004). The
$M_{500}$--$T$ relation (Fig.~\ref{f:mt}) agrees with those in Evrard
et al. (1996), Ettori et al. (2004) and Vikhlinin et al (2006). 
As an unbiased sample, the $M_{500}$--$T$ of the REFLEX-DXL sample
shows slightly higher normalization than the local relations found in
Finoguenov et al. (2001b) and Popesso et al. (2005), and the local
relation derived in Arnaud et al (2005) for relaxed clusters also
based on the XMM-Newton temperature profiles. Otherwise, there is no
obvious additional evolution of the normalization of the $M$--$T$
relation after the redshift evolution correction. This was also found
for the other distant clusters in Maughan et al. (2003) and Ettori et
al. (2003).

As shown in Fig.~\ref{f:fg}, the gas approximately follows DM at radii
above $r_{2500}$ for most of the REFLEX-DXL clusters. The gas mass can
thus be used as a measure of the total mass. We fixed the slope
parameter to 1.8 as also used in Castillo-Morales \& Schindler (2003)
and Borgani et al. (2004) and fitted the normalization. The $M_{\rm
gas,500}$--$T$ relation for the REFLEX-DXL clusters (see
Table~\ref{t:mtx_lite} and Fig.~\ref{f:mgt500}) is in good agreement
with the relation in Mohr et al. (1999) also using an X-ray flux
limited cluster sample. Our result also agrees with the recent result
in Castillo-Morales \& Schindler (2003). Recent simulations also
indicate a strong $M_{\rm gas}$--$T$ scaling relation (e.g. Borgani et
al. 2004). The normalization of the $M_{\rm gas,500}$--$T$ relations
is slightly higher for the observations than for the simulations
(e.g. Borgani et al. 2004).

The X-ray luminosity is a key parameter among the fundamental cluster
properties including also mass, temperature, and velocity
dispersion. Excluding cooling cores ($\sim 0.1 h_{50}^{-1}$~Mpc),
Markevitch (1998) reduced the scatter in the $L^{\rm bol}_{\rm
X}$--$T$ relation. As listed in Table~\ref{t:global}, we list the
bolometric X-ray luminosity of the REFLEX-DXL clusters including and
excluding the cluster cores, $r< 0.1 r_{500}$, as is used in many
studies (e.g. Markevitch 1998; Zhang 2001). The scatter of the
$L$--$T$ relation is reduced by 15\% excluding cooling cores.  About
8--33\% of the luminosity is contributed by the cluster cores ($< 0.1
r_{500}$) for the REFLEX-DXL clusters. Therefore, the normalization of
the $L$--$T$ relation excluding cooling cores is also reduced by 10\%
for the REFLEX-DXL clusters. As listed in Table~\ref{t:mtx_lite} and
shown in Figs.~\ref{f:l0124t}--\ref{f:lbolt}, we fixed the slope
parameters and fitted the normalizaton for the REFLEX-DXL sample after
the redshift evolution correction. Within the observational scatter,
the $L^{\rm bol}_{\rm X}$--$T$ relation for the REFLEX-DXL sample
(Fig.~\ref{f:lbolt}) agrees with the relation in Reiprich \&
B\"ohringer (2002) also as an unbiased sample, and the relations in
Arnaud \& Evrard (1999) and Markevitch (1998).  An alternative
redshift evolution in the $L$--$T$ relation is described in Kotov \&
Vikhlinin (2005) yielding $L\propto T^{2.64}(1+z)^{1.8}$. The result
here agrees with theirs when the alternative redshift evolution is
adopted.

In Fig.~\ref{f:lm} and Fig.~\ref{f:lbolm}, we show the $L$--$M$
relations (Table~\ref{t:mtx_lite}) of the REFLEX-DXL sample. The slope
parameter is fixed to 1.3, as derived in Popesso et al. (2005), in the
fitting procedure. The best fit of the normalization of the $L_{\rm
X}^{\rm 0.1-2.4 keV}$--$M$ relation for the REFLEX-DXL sample agrees
with the best fits in Reiprich \& B\"ohringer (2002) and Popesso et
al. (2005).

In Fig.~\ref{f:fgratio}, we show the gas mass fraction, $f_{\rm
gas,500}$, and gas mass fraction ratio, $f_{\rm gas,500}/f_{\rm
gas,2500}$, as a function of cluster temperature. We found a weak
evidence that the gas mass fraction increases with the cluster
temperature, $f_{\rm gas,500} \propto T^{0.5\pm 0.3}$. This agrees
with the scaling, $f_{\rm gas,500} \propto T^{0.34\pm 0.22}$, found in
Mohr et al (1999). The ratio of the gas mass fractions at larger and
smaller radii (e.g. $r_{500}$ and $r_{2500}$) can be used to
characterize the extent of gas relative to DM.  When this ratio is
greater than 1, the gas is more extended than DM (Reiprich 2001). For
the REFLEX-DXL clusters, the gas mass ratios (Fig.~\ref{f:fgratio})
show that gas is more extended than DM in the cluster inner region but
follows DM better in the outer region. Fig.~\ref{f:fgratio} indicates
a small trend that the gas is more extended than DM in massive
clusters. This indicates that the ICM is less influenced by
non-gravitational effects and that the energy input is less important
in the outer region for such massive clusters.  This is also indicated
by simulations (Rowley et al. 2004).

\subsection{Intrinsic scatter of the scaling relations}
\label{s:scatter}

The key to extracting cosmological parameters from the number density
of galaxy clusters is a correct understanding of the mass--observable
scaling relations and their intrinsic scatter.  The scatter of the
mass--observable scaling relation describes how well the observable
can be used as an estimator of the total mass.  The correlative
scatter ($\overline{\sigma}_{\rm cor}$) includes the intrinsic scatter
($\overline{\sigma}_{\rm int}$) and observational scatter
($\overline{\sigma}_{\rm obs}$). We investigated the logarithmic
intrinsic scatter of the scaling relations, $L$--$T$, $L$--$M$,
$M$--$T$, and $M_{\rm gas}$--$T$, of the REFLEX-DXL sample.

\begin{figure*}
\begin{center}
\includegraphics[angle=0,width=8.5cm]{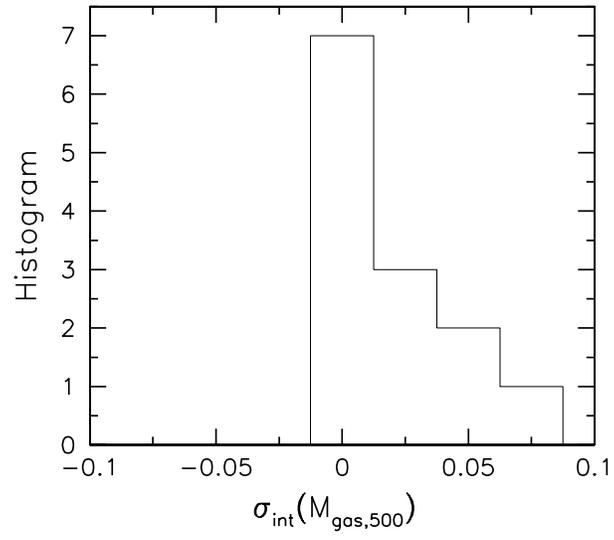}
\end{center}
\caption{Histogram of the intrinsic scatter for $\lg(M_{\rm gas,500})$
of the $M_{\rm gas,500}$--$T$ relation for the REFLEX-DXL clusters.
\label{f:scaint}
}
\end{figure*}

The observational scatter $\overline{\sigma}_{\rm obs}$ is the average
of the estimated observational uncertainties $\sigma_{\rm obs}$. The
correlative scatter $\overline{\sigma}_{\rm cor}$ is the average of
the deviation $\sigma_{\rm cor}$ of the observational data points from
the best fit of the scaling relation. Assuming that $\sigma_{\rm obs}$
and $\sigma_{\rm cor}$ are not correlated, we apply a Gaussian
statistical addition of the two effects to compute the logarithmic
intrinsic scatter $\overline{\sigma}_{\rm int}$, the average of
$\sigma_{\rm int}=\sqrt{\sigma_{\rm cor}^2-\sigma_{\rm obs}^2}$~. The
reason that we used the average instead of the mean to calculate the
scatter lies as following. We investigated the actual distributions of
the observational scatter, correlative scatter and logarithmic
intrinsic scatter. We found that in most cases the distribution
deviates from a Gaussian.  Fig.~\ref{f:scaint} shows for example the
distribution of the logarithmic intrinsic scatter of $M_{\rm gas,500}$
for the $M_{\rm gas,500}$--$T$ relation. The merger clusters
(e.g. RXCJ0516.7$-$5430 and RXCJ0658.5$-$5556) often fall into the
right tail of the histogram. The asymmetry of the histogram is most
probably due to the variety of cluster morphologies, e.g. mergers,
which can produce significant deviation from the mean. It could also
be due to the fact that the REFLEX-DXL sample does not contain enough
members to give a pronounced Gaussian statistics. We thus use the
average of the histogram to derive the logarithmic intrinsic scatter
for the whole sample, which is listed in Table~\ref{t:scaint} for the
scaling relations.

The mass--observable scaling relation which shows the least scatter
provides the best prediction for the total mass or gas mass. We find
here that the temperature is the best estimator. However, we have to
recall that the large uncertainty in the mass estimate comes from the
uncertainty in the temperature distribution and therefore part of the
reason of the small scatter originates from the fact that the
uncertainties are correlated. This is not true for the determination
of the gas mass which does hardly depend on the temperature
measurement. Therefore the $M_{\rm gas,500}$--$T$ relation, for which
we found $\overline{\sigma}_{\rm cor}=0.16$ for $\lg(M_{\rm gas})$ for
the REFLEX-DXL sample, is the most remarkably tight correlation.

The scatter of the scaling relations of the REFLEX-DXL sample confirms
the recent studies in observations (e.g. Reiprich \& B\"ohringer 2002)
and simulations (e.g. Borgani et al. 2004). For example, we obtained
the correlative scatter of (0.19,0.20) for ($\lg(L)$,$\lg(M)$) in the
$L$--$M_{500}$ relation and confirm the recent studies in observations
(e.g. Reiprich \& B\"ohringer 2002) and simulations (e.g. Borgani et
al. 2004).

\section{Discussion}
\label{s:discu}

\subsection{Gas profiles in the outskirts}
\label{s:infall}

The generally adopted $\beta$-model ($\beta=2/3$) gives $n_{\rm e}
\propto r^{-2}$. Mass in-falling becomes significant in the outskirts
and thus makes the electron density profile steeper.  Vikhlinin et
al. (1999) found a mild trend for $\beta$ to increase as a function of
cluster temperature, which gives $\beta \sim 0.80$ and $n_{\rm e}
\propto r^{-2.4}$ for clusters around 10~keV.  Bahcall (1999) also
found that the electron number density scales as $n_{\rm e} \propto
r^{-2.4}$ at large radii. We confirm their conclusion that $n_{\rm e}
\propto r^{-2.42}$ at $r>2^{\prime}$ for the REFLEX-DXL clusters. Due
to the gradual change in the slope, one should be cautious to use a
single slope double-$\beta$ model which might introduce a systematic
error in the cluster mass measurements (as also described
e.g. in Horner 2001).

\subsection{Validity of the spherical symmetry and hydrostatic equilibrium}

The total mass can be underestimated due to the assumption of
spherical symmetry for elongated clusters (Castillo-Morales \&
Schindler 2003). However, a low surface brightness extension is often
difficult to subtract correctly because of its less significant
boundary relative to the surroundings. In the REFLEX-DXL sample,
RXCJ0516.7$-$5430 is such an extreme case. A compression of the photon
distribution extends from the cluster center to the north and a low
surface brightness extends to $r_{500}$ south of the cluster
center. We investigated this cluster and found the global measurements
such as $M_{500}$ are insensitive to the inclusion or removal of the
region of the low surface brightness extension. The difference of the
surface brightness profiles including and excluding the low surface
brightness extension is within the 1-$\sigma$ error bar of the surface
brightness.

When X-ray images display a pronounced elongated and distorted cluster
morphology, the cluster central position cannot be determined
unambiguously. For example, RXCJ0014.3$-$3022, RXCJ0516.7$-$5430,
RXCJ0528.9$-$3927, RXCJ0658.5$-$5556 and RXCJ1131.9$-$1955 are such
cases in the REFLEX-DXL sample. For the case RXCJ0528.9$-$3927, we
found that the measurements still agree with each other at the
1-$\sigma$ confidence level using the cluster center of the main
component and the center of mass of the two components, respectively.

Clusters showing significant merger features are still dynamically
young. These merger features can invalidate the hypothesis of
spherical symmetry and hydrostatic equilibrium. For example,
RXCJ0658.5$-$5556 is one of the most spectacular merger clusters.  We
checked the substructure excision method in this work by comparing the
results including and excluding the substructure in
RXCJ0658.5$-$5556. We found that the global measurements are
insensitive to the method excising substructure within the 1-$\sigma$
observational error.

\subsection{RXCJ0658.5$-$5556 and the scaling relations}

It is worthy to take a closer look, how clusters of particular
morphological type affect the scaling relations. We find that the
pronounced merger clusters (RXCJ0014.3$-$3022, RXCJ0516.7$-$5430,
RXCJ2337.6$+$0016 and RXCJ0658.5$-$5556), which also show an excess in
the central entropies and/or large gas mass fractions, introduce a
significant broadening of the scatter in the scaling
relations. Therefore the scatter has to be considered with caution
since it could partially be an artificial effect of the invalidity of
hydrostatic equilibrium. A particular case is the extremely hot
cluster, RXCJ0658.5$-$5556. For example, excluding RXCJ0658.5$-$5556
the $M_{\rm gas,500}$--$T$ relation provides reduced scatter,
$\overline{\sigma}_{\rm cor}=0.11$ and $\overline{\sigma}_{\rm
cor}=0.06$ for $M_{\rm gas,500}$ and $T$, respectively.

RXCJ0658.5$-$5556 has a very large weight on the slope of the scaling
relations due to its extreme location on the parameter scale. It is
also one of the merger systems with large uncertainties of the
observational parameters. Therefore we have to be careful with the
slope fitting when such merger clusters are involved. As an extreme
case, the $S$--$T$ relation of all the REFLEX-DXL clusters gives a
slope twice higher than the published empirical slope. In this work,
we either fitted the slope parameters excluding merger clusters
(e.g. $S$--$T$) or fixed the slope parameters to those in the
published papers (e.g. $M$--$T$).

Merger clusters might also affect the determination of the
normalization of the scaling relations. Excluding RXCJ0658.5$-$5556,
the normalization of the scaling relations of $M_{500}$--$T$,
$M_{\rm gas,500}$--$T$, and $L$--$T$ will be reduced by 5\%, 5\%, and 4\%,
respectively.  RXCJ0658.5$-$5556 lays above the normalized relations
by a factor of 1.9,1.5, and 1.1 for $M_{500}$--$T$,
$M_{\rm gas,500}$--$T$, and $L$--$T$, respectively.

\subsection{Dense core or cool gas}

A dense gaseous cluster core, as observed in the CCCs, does not
necessarily require a cusp of the DM distribution as described by the
extended-NFW model. Alternatively, a sufficiently cooler central
temperature also results in a dense core without a cuspy DM profile
which is demonstrated as follows. Restricting the analysis to
$r<0.5^{\prime}$, we use the pronounced CCC, RXCJ0307.0$-$2840, to
illustrate the total mass distribution in the dense and cool gaseous
region in the cluster. In general, no cool gas has been observed
showing a central temperature lower than half of the mean temperature.
We thus assume an extreme temperature drop to 1/3 of the observed
temperature of the inner most bin towards the center. Assuming
hydrostatic equilibrium, we derived a relatively low mass
concentration in the cluster center with the steep central gas density
distribution. We can not easily distinguish between the above two
cases using the current observations.

\subsection{Additional physical processes}

We observed a deviation around the self-similar model in the central
region in the scaled profiles of the temperature, surface brightness,
entropy, total mass and gas mass. This is most probably the effect of
different physical processes rather than simply being statistical
fluctuations in the measurements (Zhang et al. 2004b, 2005b;
Finoguenov et al. 2005). Many studies (e.g. Markevitch et al. 2002;
Randall et al. 2002; Finoguenov et al. 2005) show that the X-ray
property estimates in the center can be biased by phenomena such as
ghost cavities, bubbles, shock and cold fronts, that may somehow
invalidate the hydrostatic equilibrium hypothesis and the assumption
of homogeneous temperature and density distributions. Complex
dynamical interactions with AGN activities have been indicated by the
coincidence of CCCs and radio sources (Clarke et al. 2005). M87 shows
an example to test the effect of AGN interaction on the X-ray luminosity
and also multi-temperature structure (Matsushita et al. 2002).

After subtraction of the effect of measurement uncertainties, the
remaining intrinsic scatter in the scaling relations is a signature in
variations of cluster structure and ICM processes. This also includes
merging clusters as an extreme case (Zhang et al. 2004b). Systematic
studies of X-ray mergers have been done in observations using a series
of cluster samples (e.g. Schuecker et al. 2001) and simulations
(e.g. Schindler \& Mueller 1993). Such a detailed study of the
REFLEX-DXL clusters using a 2-dimensional approach can be found in
Finoguenov et al. (2005).

\section{Summary and conclusions}
\label{s:conclusion}

X-ray luminous (massive) clusters are used in a variety of ways to
perform both cosmological and astrophysical studies. We selected an
unbiased, almost volume-limited sample, the REFLEX-DXL cluster sample,
from the REFLEX survey. We performed a systematic analysis to measure
the X-ray observables based on high quality XMM-Newton observations,
and investigated various X-ray properties and the scaling relations of
the REFLEX-DXL cluster sample. We summarize two main conclusions as
follows.

\bigskip

(i) An almost self-similar behaviour of the scaled profiles of X-ray
properties, such as temperature, surface brightness, entropy,
gravitational mass, and gas mass has been found above
0.2--0.3~$r_{500}$ for the REFLEX-DXL sample.

\begin{figure*}
\begin{center}
\includegraphics[width=6.5cm,angle=270]{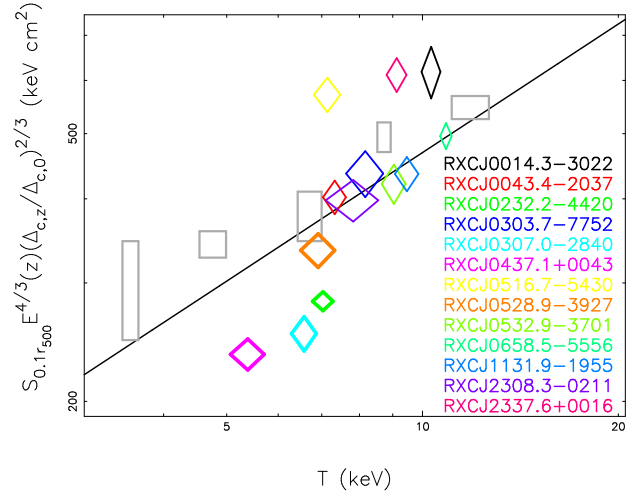}
\end{center}
\caption{Redshift corrected entropy (diamonds) at $0.1 r_{\rm 500}$
vs. $T$ for the REFLEX-DXL clusters. Four clusters showing relatively
cool cluster cores are in thick lines. Nearby clusters of Ponman et
al. (2003, boxes) are shown for comparison. The solid line denotes the
best fits ($S \propto T^{0.63 \pm 0.15}$). {\it See the electronic
edition of the Journal for a colour version.}
\label{f:cores}
}
\end{figure*}

\begin{figure*}
\begin{center}
\includegraphics[width=6.5cm,angle=270]{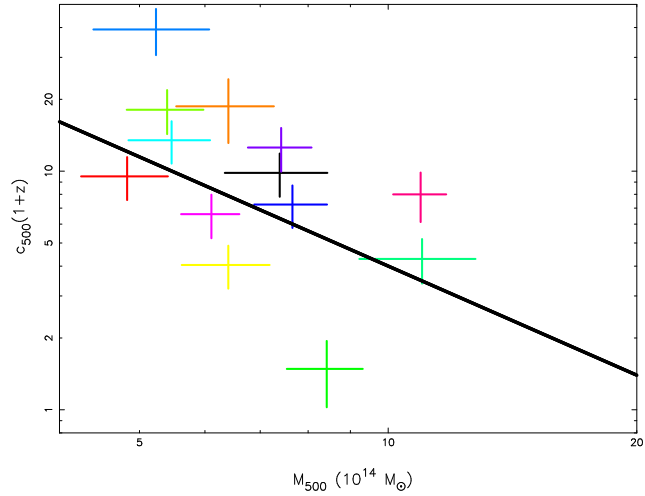}
\end{center}
\caption{The concentration parameter of the extended-NFW model versus
mass for the REFLEX-DXL sample. The line denotes the best fit of the
REFLEX-DXL sample. The colours have the same meaning as those in
Fig.~\ref{f:cores}. \label{f:cmass} }
\end{figure*}

\begin{figure*}
\begin{center}
\includegraphics[width=6.5cm,angle=270]{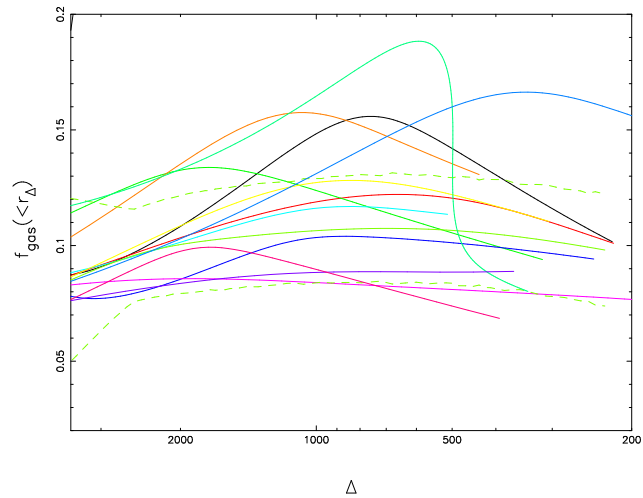}
\end{center}
\caption{Gas mass fraction as a function of density contrast for the
REFLEX-DXL sample. The 1-$\sigma$ error corridor (dashed curves)
is displayed for RXCJ0532.9$-$3701 only. The colours have the same
meaning as those in Fig.~\ref{f:cores}. \label{f:fg} }
\end{figure*}

\begin{figure*}
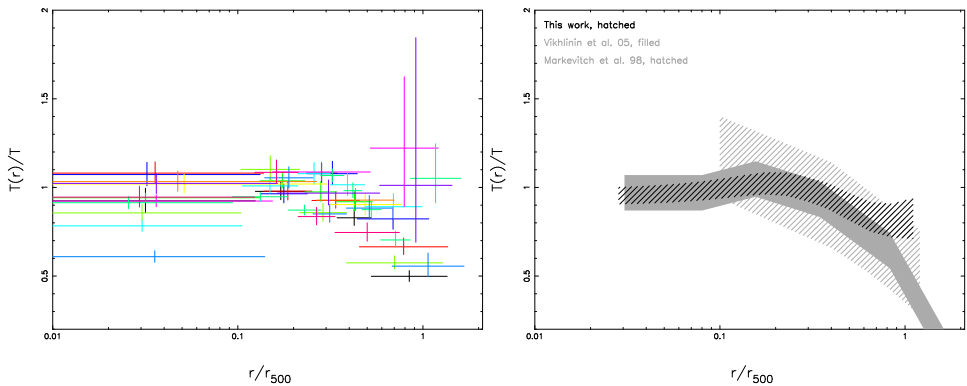

\begin{center}
\includegraphics[width=5cm,angle=270]{3650f8a.ps}
\includegraphics[width=5cm,angle=270]{3650f8b.ps}
\end{center}
\caption{{\it Left:} Scaled radial
temperature profiles. The colours have the same meaning as those in
Fig.~\ref{f:cores}. {\it Right:} An average temperature profile of the
REFLEX-DXL clusters compared to the temperature profile ranges in
Markevitch et al. (1998, grey, hatched) and Vikhlinin et al. (2005,
grey, filled).
\label{f:scalet}
}
\end{figure*}

\begin{figure*}
\begin{center}
\includegraphics[width=6.5cm,angle=270]{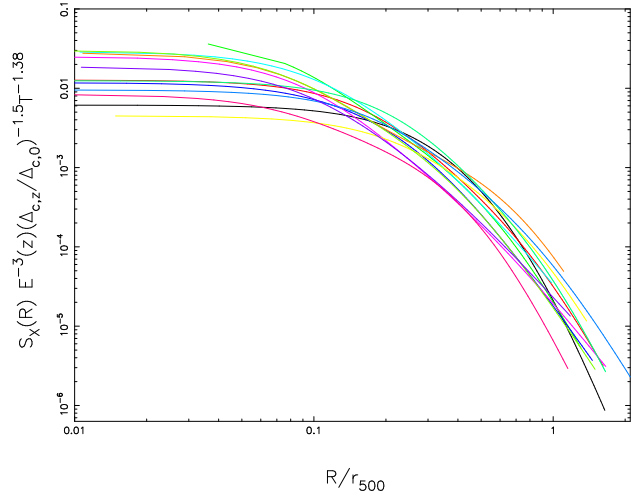}
\end{center}
\caption{Surface brightness profile fits scaled according to the
empirical scaling, $S_{\rm X} \propto T^{1.38}$ given in Arnaud et
al. (2002). The colours have the same meaning as those in
Fig.~\ref{f:cores}.
\label{f:scalesx}
}
\end{figure*}

\begin{figure*}
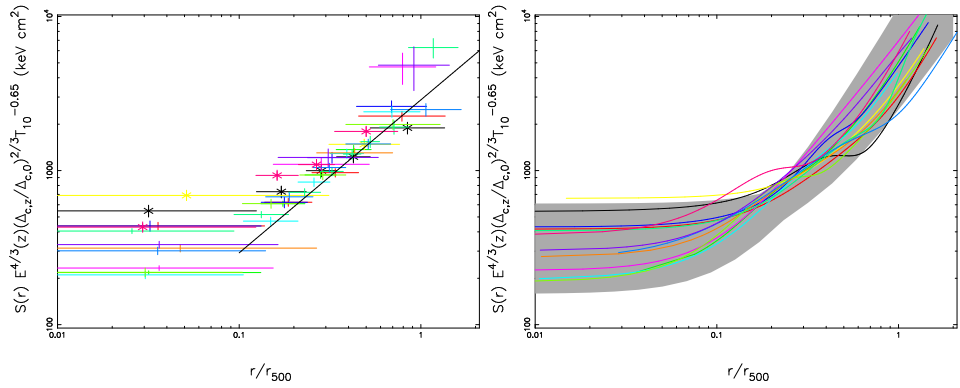

\begin{center}
\includegraphics[width=5cm,angle=270]{3650f10a.ps}
\includegraphics[width=5cm,angle=270]{3650f10b.ps}
\end{center}
\caption{{\it Left:} Scaled entropy profiles and the best fit of the
$r>0.1 r_{500}$ region. The merger clusters (RXCJ0014.3$-$3022,
RXCJ0516.7$-$5430 and RXCJ2337.6$+$0016) are shown with asterisk
symbols. $T_{10}$ denotes temperatures in units of 10~keV. {\it
Right:} Scaled entropy profile fits of the REFLEX-DXL sample compared
to the Birmingham-CfA clusters (grey shadow) in a temperature range of
6--20~keV. The colours have the same meaning as those in
Fig.~\ref{f:cores}.
\label{f:en}
}
\end{figure*}

\begin{figure*}
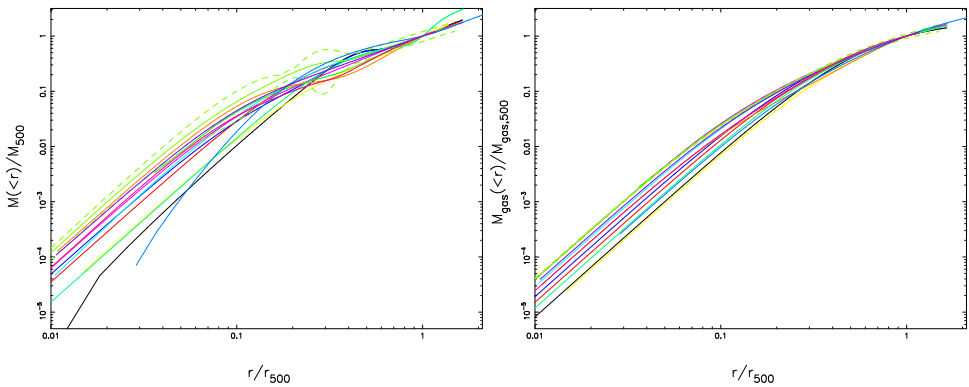

\begin{center}
\includegraphics[width=5cm,angle=270]{3650f11a.ps}
\includegraphics[width=5cm,angle=270]{3650f11b.ps}
\end{center}
\caption{Scaled total mass profiles (left) and gas mass profiles
(right) for the REFLEX-DXL sample, with the 1-$\sigma$ error corridor
(dashed) displayed for RXCJ0532.9$-$3701. The colours have the same
meaning as those in Fig.~\ref{f:cores}.
\label{f:scaleym}
}
\end{figure*}

\begin{figure*}
\begin{center}
\includegraphics[angle=270,width=8.5cm]{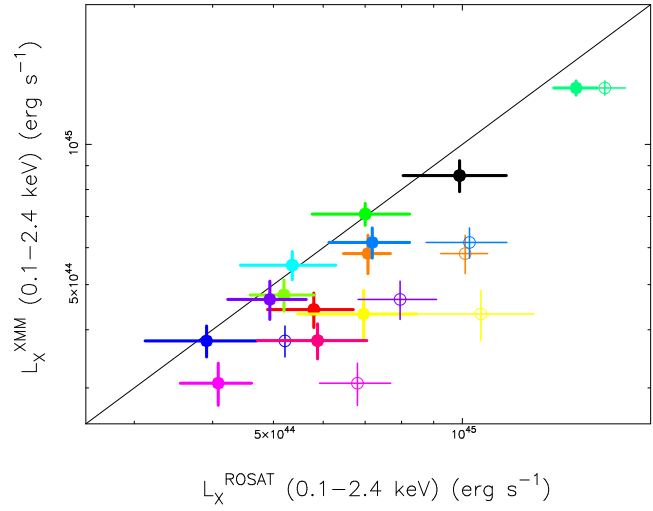}
\end{center}
\caption{A comparison of the XMM-Newton luminosity and ROSAT
luminosity taken from B\"ohringer et al. (2004). Filled and open
circles correspond to the substructure and point-like source
subtracted and un-subtracted ROSAT luminosity, respectively. The
point-like sources and substructures are detected using the XMM-Newton
observations. The colours have the same meaning as those in
Fig.~\ref{f:cores}. \label{f:lxmmrosat}}
\end{figure*}

\begin{figure*}
\begin{center}
\includegraphics[angle=270,width=8.5cm]{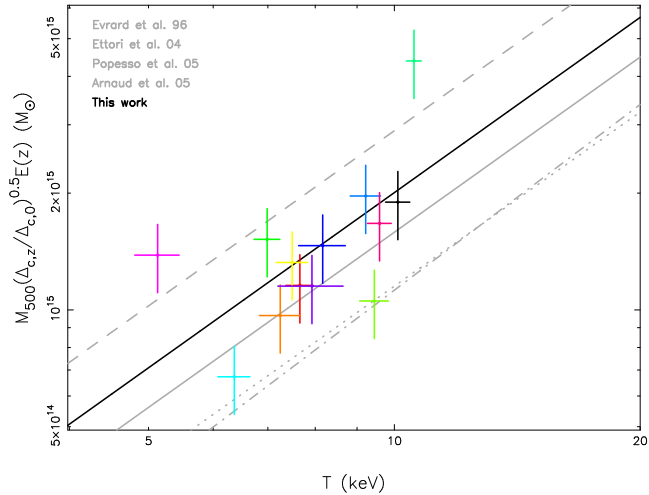}
\end{center}
\caption{Mass vs. temperature for the REFLEX-DXL sample and the best
fit power law (black).  The grey lines denote the best fits in Evrard
et al. (1996, solid), Ettori et al. (2004, dashed), Popesso et al.
(2005, dash-dotted) and Arnaud et al. (2005, dotted). The colours have
the same meaning as those in Fig.~\ref{f:cores}.
\label{f:mt}}
\end{figure*}

\begin{figure*}
\begin{center}
\includegraphics[angle=270,width=8.5cm]{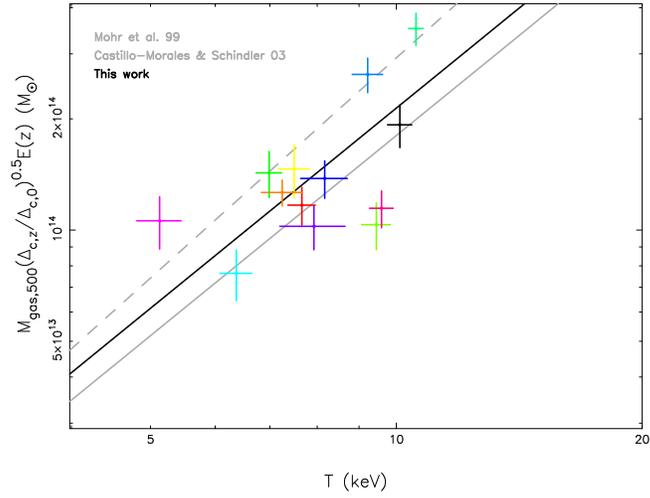}
\end{center}
\caption{Gas mass vs. cluster temperature and the best fit power law
(black).  The grey lines denote the best fits in Mohr et al. (1999,
dashed) and Castillo-Morales \& Schindler (2003, solid). The colours
have the same meaning as those in Fig.~\ref{f:cores}.
\label{f:mgt500}}
\end{figure*}

\begin{figure*}
\begin{center}
\includegraphics[angle=270,width=8.5cm]{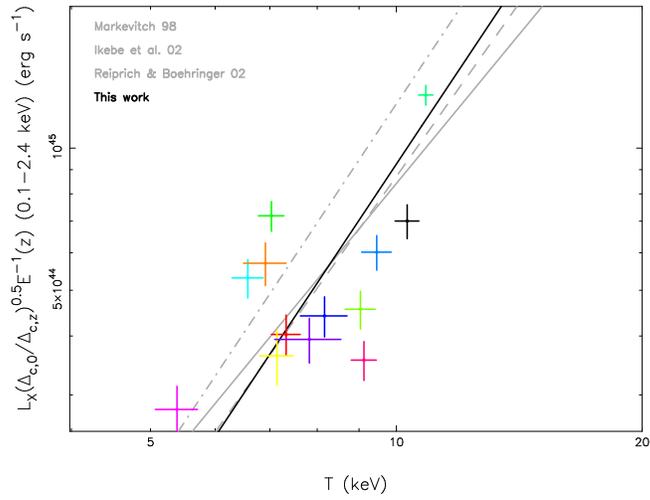}
\end{center}
\caption{X-ray luminosity in the 0.1--2.4~keV band vs. cluster
temperature and the best fit power law (black). The grey lines denote the best
fits in Markevitch (1998, solid), Ikebe et al. (2002, dashed), and
Reiprich \& B\"ohringer (2002, dash-dotted). The colours have the same meaning as those in Fig.~\ref{f:cores}.
\label{f:l0124t}
}
\end{figure*}

\begin{figure*}
\begin{center}
\includegraphics[angle=270,width=8.5cm]{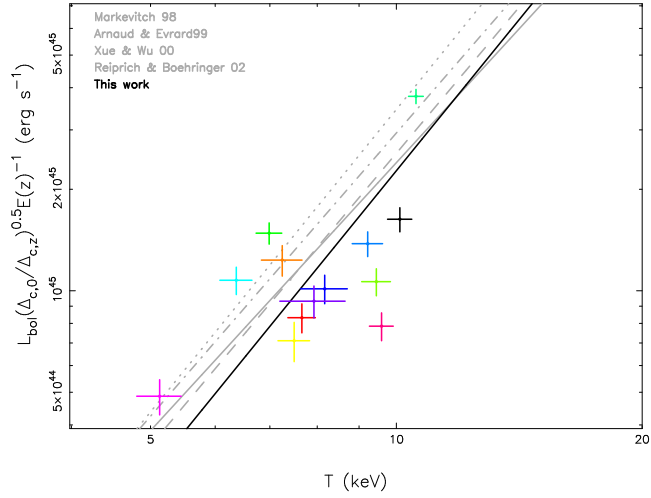}
\end{center}
\caption{X-ray bolometric luminosity vs. cluster temperature and the
best fit power law (black). The grey lines denote the best fits in Markevitch
(1998, solid), Arnaud \& Evrard (1999, dashed), Xue \& Wu (2000,
dash-dotted), and Reiprich \& B\"ohringer (2002, dotted). The colours
have the same meaning as those in Fig.~\ref{f:cores}.
\label{f:lbolt}
}
\end{figure*}

\begin{figure*}
\begin{center}
\includegraphics[angle=270,width=8.5cm]{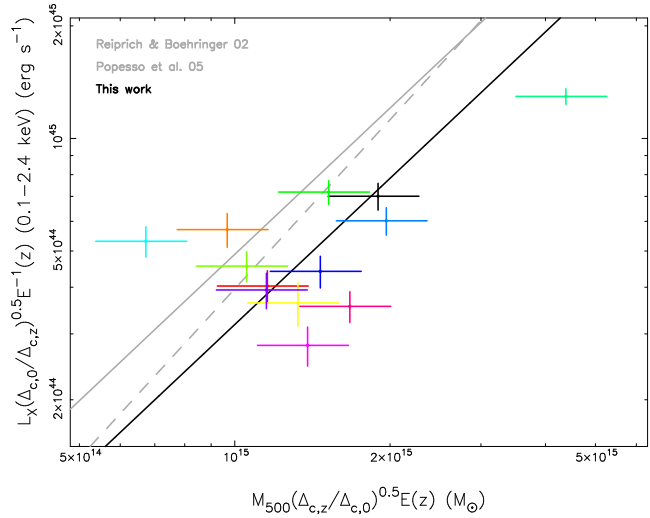}
\end{center}
\caption{X-ray luminosity in the 0.1--2.4~keV band vs. total mass and
the best fit power law (black). The grey lines denote the best fits in
Reiprich \& B\"ohringer (2002; dashed) and Popesso et al.(2005,
solid). The colours have the same meaning as those in
Fig.~\ref{f:cores}.
\label{f:lm}}
\end{figure*}

\begin{figure*}
\begin{center}
\includegraphics[angle=270,width=8.5cm]{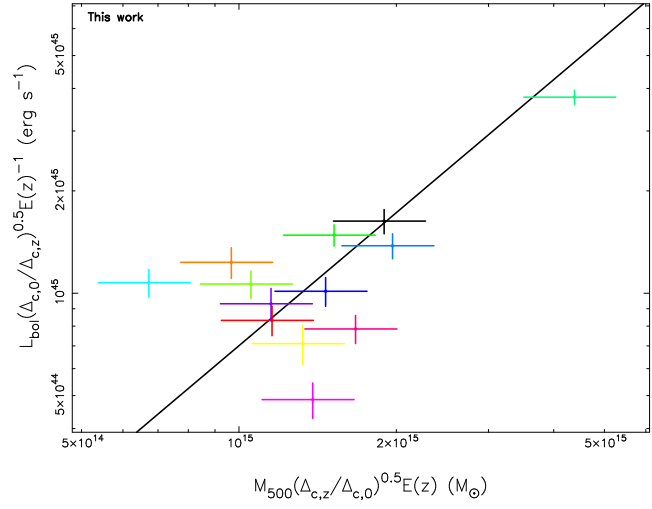}
\end{center}
\caption{X-ray bolometric luminosity vs. cluster total mass and the
best fit power law (black). The colours have the same meaning as those in
Fig.~\ref{f:cores}.
\label{f:lbolm}
}
\end{figure*}

\begin{figure*}
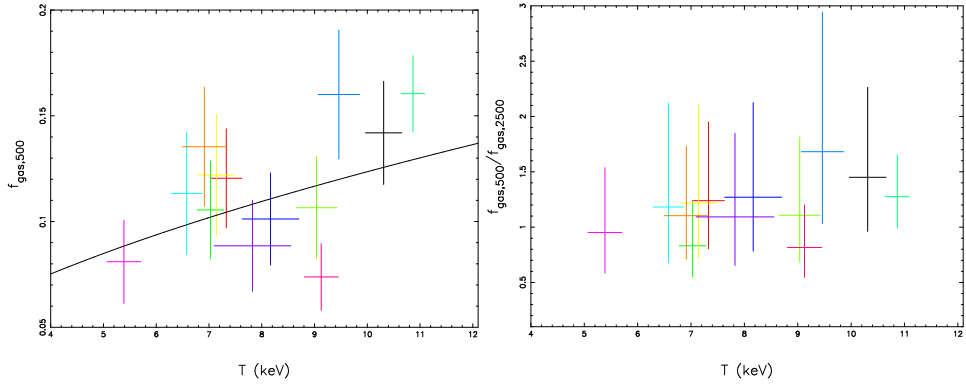

\begin{center}
\includegraphics[width=5cm,angle=270]{3650f19a.ps}
\includegraphics[width=5cm,angle=270]{3650f19b.ps}
\end{center}
\caption{Gas mass fraction $f_{\rm gas,500}$ (left) and
Gas mass fraction ratio $f_{\rm gas,500}/f_{\rm gas,2500}$ (right) as
a function of cluster temperature. The best fit is $f_{\rm gas,500}
\propto T^{0.5 \pm 0.3}$. The colours have the same meaning as those
in Fig.~\ref{f:cores}.
\label{f:fgratio}
}
\end{figure*}

\begin{itemize}

\item 
The average global metallicity is $0.24\pm 0.03 Z_{\odot}$
for the whole sample. No significant evolution was found up to $z\sim
0.3$ in the metallicity comparing the REFLEX-DXL sample to the nearby
galaxy cluster samples. This agrees with the results in Allen \&
Fabian (1998), Tozzi et al. (2003), Chen et al. (2003), De Grandi et
al. (2004) and Pointecouteau et al. (2004). The results fit into the
scenario showing no significant evolution of the iron abundance up to
$z\sim 1.1$.

\item
Based on the XMM-Newton observations, we obtained an average
temperature profile of the REFLEX-DXL clusters, which agrees with the
previous studies within the observational dispersion. Markevitch et
al. (1998) found a steep temperature drop beyond an isothermal center
based on ASCA data. De Grandi \& Molendi (2002) derived a universal
temperature profile which shows a similar decline using BeppoSAX
observations. Based on high resolution Chandra observations, Vikhlinin
et al. (2005) and Piffaretti et al. (2005) confirmed the previous
studies within the observational dispersion, where they found a more
pronounced drop outside of 0.2--0.3~$r_{\rm 200}$ in the mean of the
universal temperature profile. Additionally, Borgani et al. (2004)
reproduced a similar temperature profile in their simulations as found
by Vikhlinin et al. (2005). For the REFLEX-DXL sample, we found a
universal temperature profile with a peak around
0.2--0.3~$r_{500}$. We observed an almost constant value up to
0.3$r_{500}$ and a very mild decrease to 80\% of the peak value at $r>
0.3 r_{500}$ for the average temperature radial profile of the
REFLEX-DXL clusters. However, three clusters (RXCJ0437.1$+$0043,
RXCJ0658.5$+$5556, and RXCJ2308.3$+$0211) show temperatures rising
with relatively large error bars in their temperature profiles. The
average temperature profile declines to 50\% of the peak value at
radii above $0.3 r_{500}$ when these three clusters are excluded. No
significant detection of cool gas has been observed showing a central
temperature lower than half of the mean temperature.

\item
We determined the XMM-Newton surface brightness profiles of
the REFLEX-DXL clusters up to at least $r_{500}$. We observe steeper
profiles at large radii than what is generally obtained for the
$\beta$-models with $\beta \sim 2/3$. The surface brightness profiles
of most REFLEX-DXL clusters show a flat core populating a broad range
of values up to 0.2~$r_{500}$. The total luminosity of the cluster
core ($< 0.1 r_{500}$) accounts for about 8--33\% cluster
luminosity. For the clusters showing cooling cores, no well-defined
constant central density was observed with the XMM-Newton
resolution. However, the surface brightness profiles are quite
self-similar at $r>0.2 r_{500}$ for the REFLEX-DXL sample.

\item 
We performed the redshift evolution correction to the
entropy profiles at $0.1 r_{\rm 500}$ for the REFLEX-DXL clusters and
obtained consistency with those for the nearby clusters in Ponman et
al. (2003) (see Fig.~\ref{f:cores} and Fig.~\ref{f:en}) within the
observational dispersion. However, four clusters showing relatively
cool cores introduce a significant scatter in terms of a lower 
($\sim 7\%$) normalization of the $S$--$T$ relation, and three merger
clusters introduce some scatter in terms of a higher ($\sim 10\%$)
normalization in the $S$--$T$ relation. Excluding these 7 clusters,
the REFLEX-DXL sample shows an empirical scaling,
$T^{0.63\pm0.15}$. As shown in Fig.~\ref{f:en}, the entropy profiles
at $r>0.1r_{500}$ show a similar slope ($S \propto r^{0.99\pm 0.06}$)
as observed in Ettori et al. (2002b), $S\propto r^{0.97}$. At the
1-$\sigma$ confidence level, this scaling is shallower than the
prediction of the spherical accretion shock model, $S\propto r^{1.1}$
(e.g. Tozzi \& Norman 2001; Kay 2004). We found that merger clusters
show high central entropies, and the relatively relaxed clusters show
lower central entropies. Therefore the central entropy can be used as
one means to distinguish the cluster dynamic state. For example, the
observational deviation of the ($S$,$T$) data pair from the $S$--$T$
scaling relation for the individual cluster can be used to distinguish
the relaxation stage. This may be particularly useful to
characterize cluster merging along the line-of-sight, where the merger
morphology is less obvious in the projection on the sky.

\item 
We found a self-similar gravitational mass distribution for
the REFLEX-DXL sample at radii above 0.2--0.3~$ r_{\rm 500}$. The
precise gas density and temperature radial profiles provide a
detailed diagnostics of the cluster structure and yield reliable
determinations of the total mass and gas mass fraction. In the
outskirts, the observational density distribution provides an average
of $\rho(r) \propto r^{-2.42}$ for the REFLEX-DXL sample.

\end{itemize}

\bigskip

(ii) The scaling relations of the REFLEX-DXL sample at $z \sim 0.3$
agree with the scaling relations of the nearby samples within the
observational dispersion after the redshift evolution correction.

\begin{itemize}

\item
Since the cluster temperatures of the REFLEX-DXL sample are
in a narrow temperature range, the whole sample provides a good means
to constrain the normalization of the scaling relations (see
Table~\ref{t:mtx_lite}) and to study their intrinsic scatter (see
Table~\ref{t:scaint}) at the high mass end.

\item 
The results for the scaling relations of the REFLEX-DXL
sample show good agreement compared with previous studies
(Table~\ref{t:mtx_lite}). This fits the general opinion (e.g. Maughan
et al. 2003; Arnaud 2005; Vikhlinin et al. 2006) that the evolution of
galaxy clusters up to $z \sim 1$ is well described by a self-similar
model for massive clusters.

\item
We found that the scatter of the normalization of the
scaling relations is very sensitive to the cluster morphology. For
example, the scatter of the $L$--$T$ relation is reduced by 15\%
excluding 4 REFLEX-DXL clusters showing pronounced radiative
cooling. Also the normalization of the $L$--$T$ relation excluding
those 4 clusters is reduced by 10\% for the REFLEX-DXL clusters. We
investigated the logarithmic intrinsic scatter of the scaling
relations which, for example, gives (0.19,0.20) for ($L$,$M$) in the
$L$--$M_{500}$ relation and confirms the recent studies, such as
(0.29,0.22) in Reiprich \& B\"ohringer (2002).

\end{itemize}

\begin{acknowledgements}

The XMM-Newton project is supported by the Bundesministerium f\"ur
Bildung und Forschung, Deutschen Zentrum f\"ur Luft und Raumfahrt
(BMBF/DLR), the Max-Planck Society and the Haidenhaim-Stiftung. We
acknowledge Jacqueline Bergeron, PI of the XMM-Newton observation of
the CDFS, and Martin Turner, PI of the XMM-Newton observation of
RXJ0658.5-5556.  We acknowledge the anonymous referee for providing
lots of comments to improve the work. We acknowledge Alexey Vikhlinin,
Naomi Ota, Michael Freyberg, Ulrich G. Briel, Gabriel Pratt, Takaya
Ohashi and Rasmus Voss for providing useful
suggestions. Y.Y.Z. acknowledges support under the XMM-Newton grant
No.\,NNG\,04\,GF\,68\,G. A.F. acknowledges support from BMBF/DLR under
grant No.\,50\,OR\,0207 and MPG.

\end{acknowledgements}


\clearpage

\begin{table*} { \begin{center} \footnotesize
{\renewcommand{\arraystretch}{1.3} \caption[]{ Properties of the
suspicious point-like sources in the cluster center.
Column~(1): cluster name; Cols.~(2,3): sky coordinates in epoch J2000
of the point-like source; Col.~(4): index of the best power law
fit; Col.~(5): X-ray luminosity of the point-like source.}
  \label{t:centpsrc}}
\begin{tabular}{ccccc}
\hline
\hline
RXCJ & \multicolumn{2}{c}{X-ray centroid} & Power law & $L_{\rm X}$ (0.1--2.4~keV) \\
     &    R.A.       &   decl.          &           & $10^{44}$~erg~s$^{-1}$ \\
\hline
0232.2$-$4420 & 02 32 18.6 & $-$44 20 48.2 & $1.77\pm0.03$ & $5.22\pm0.27$\\
0437.1$+$0043 & 04 37 09.8 & $+$00 43 48.9 & $1.82\pm0.03$ & $3.56\pm0.18$\\
0528.9$-$3927 & 05 28 52.6 & $-$39 28 16.8 & $1.70\pm0.05$ & $3.20\pm0.27$\\
1131.9$-$1955 & 11 31 54.2 & $-$19 55 39.8 & $1.75\pm0.04$ & $1.97\pm0.08$\\
2308.3$+$0211 & 23 08 21.6 & $-$02 11 29.1 & $1.67\pm0.05$ & $2.24\pm0.12$\\
2337.6$+$0016 & 23 37 35.3 & $+$00 15 52.1 & $1.80\pm0.06$ & $1.02\pm0.10$\\
\hline
\hline
  \end{tabular}
  \end{center}
\hspace*{0.3cm}{\footnotesize } }
\end{table*}

\begin{table*} { \begin{center} \footnotesize
  {\renewcommand{\arraystretch}{1.3} \caption[]{Classification of the
  dynamical state. Column(1): classification; Col.(2): cluster names.}  \label{t:dyn}}
\begin{tabular}{ll}
\hline
\hline
Classification               & RXCJ \\
\hline
Single                       & 0307.0$-$2840, 0532.9$-$3701, 2308.3$-$0211                \\
Primary with small secondary & 0232.2$-$4420, 0303.7$-$7752                               \\
Elliptical                   & 0043.4$-$2037, 0437.1$+$0043, 0516.7$-$5430, 1131.9$-$1955 \\
Offset center                & 0014.3$-$3022, 0528.9$-$3927, 0658.5$-$5556, 2337.6$+$0016 \\
Complex                      & 2011.3$-$5725                                              \\
\hline
\hline
  \end{tabular}
  \end{center}
\hspace*{0.3cm}{\footnotesize Classifications see Jones \& Forman (1992).} }
\end{table*}

\begin{table*} { \begin{center} \footnotesize
  {\renewcommand{\arraystretch}{1.3} \caption[]{ Primary
  parameters. Column~(1): cluster name; Col.~(2): optical redshift
  (B\"ohringer et al. 2004); Cols.~(3,4): sky coordinates in epoch
  J2000 of the cluster center; Col.~(5): hydrogen column density in
  units of $10^{20}{\rm cm}^{-2}$ (Dickey \& Lockman
  1990); Col.~(6):X-ray measured redshift; Col.~(7): truncation radius
  corresponding to a $S/N$ of 3 of the observational surface brightness
  profile; Cols.~(8,9): cluster global temperature and global
  metallicity in the 0.1--0.5~$r_{500}$ region; Cols.~(10,11):
  bolometric luminosity including and excluding cluster core
  ($<0.1r_{500}$).}  \label{t:global}}
  \begin{tabular}{lcccccccccc}
\hline
\hline
RXCJ& $z_{\rm opt}$ & \multicolumn{2}{c}{X-ray centroid} & $N_{\rm H}$ & $z_{\rm X-ray}$ & $r_{\rm t}$ & $T$ & $Z$ & $L_{\rm bol}^{\rm incc}$ & $L_{\rm bol}^{\rm excc}$ \\
    &               &    R.A.   &  decl.    & $10^{20}$~cm$^{-2}$ & & arcmin & keV & $Z_{\odot}$ &  \multicolumn{2}{c}{$10^{45}$~erg~s$^{-1}$}\\
\hline
0014.3$-$3022 & 0.3066  &  00  14 18.6  &  $-30$  23 15.4  &  1.60  & $0.28 \pm0.01$ & 7.61 &$10.1 \pm  0.3 $&$
0.17 \pm 0.05$ 
 & $ 2.12 \pm 0.17$ & $1.91 \pm 0.16$\\
0043.4$-$2037 & 0.2924  &  00  43 24.5  &  $-20$  37 31.2  &  1.54  & $0.28 \pm0.01$ & 7.02 &$ 7.7 \pm  0.3 $&$
0.11 \pm 0.06$ 
 & $ 1.07 \pm 0.11$ & $0.89 \pm 0.09$\\
0232.2$-$4420 & 0.2836  &  02  32 18.8  &  $-44$  20 51.9  &  2.49  & $0.27 \pm0.01$ & 6.81 &$ 7.0 \pm  0.3 $&$
0.10 \pm 0.05$ 
 & $ 1.89 \pm 0.14$ & $1.27 \pm 0.11$\\
0303.7$-$7752 & 0.2742  &  03  03 47.2  &  $-77$  52 39.0  &  8.73  & $0.26 \pm0.01$ & 7.53 &$ 8.2 \pm  0.5 $&$
0.32 \pm 0.08$ 
 & $ 1.29 \pm 0.13$ & $1.04 \pm 0.11$\\
0307.0$-$2840 & 0.2578  &  03  07 02.2  &  $-28$  39 55.2  &  1.36  & $0.24 \pm0.01$ & 4.70 &$ 6.4 \pm  0.3 $&$
0.13 \pm 0.06$ 
 & $ 1.34 \pm 0.13$ & $1.00 \pm 0.10$\\
0437.1$+$0043 & 0.2842  &  04  37 09.5  &  $+00$  43 54.5  &  8.68  & $0.31 \pm0.02$ & 7.73 &$ 5.1 \pm  0.3 $&$
0.30 \pm 0.11$ 
 & $ 0.62 \pm 0.07$ & $0.42 \pm 0.06$\\
0516.7$-$5430 & 0.2943  &  05  16 35.2  &  $-54$  30 36.8  &  6.86  & $0.28 \pm0.01$ & 6.39 &$ 7.5 \pm  0.3 $&$
0.16 \pm 0.07$ 
 & $ 0.92 \pm 0.12$ & $0.84 \pm 0.11$\\
0528.9$-$3927 & 0.2839  &  05  28 52.5  &  $-39$  28 16.7  &  2.12  & $0.26 \pm0.02$ & 5.80 &$ 7.2 \pm  0.4 $&$
0.42 \pm 0.08$ 
 & $ 1.58 \pm 0.16$ & $1.25 \pm 0.13$\\
0532.9$-$3701 & 0.2747  &  05  32 55.9  &  $-37$  01 34.5  &  2.90  & $0.27 \pm0.01$ & 7.01 &$ 9.5 \pm  0.4 $&$
0.32 \pm 0.07$ 
 & $ 1.35 \pm 0.12$ & $0.95 \pm 0.10$\\
0658.5$-$5556 & 0.2965  &  06  58 30.2  &  $-55$  56 33.7  &  6.53  & $0.29 \pm0.01$ & 8.88 &$10.6 \pm  0.2 $&$
0.26 \pm 0.03$ 
 & $ 4.87 \pm 0.24$ & $4.21 \pm 0.21$\\
1131.9$-$1955 & 0.3075  &  11  31 54.7  &  $-19$  55 40.5  &  4.50  & $0.33 \pm0.03$ & 9.00 &$ 9.2 \pm  0.4 $&$
0.27 \pm 0.05$ 
 & $ 1.80 \pm 0.15$ & $1.58 \pm 0.14$\\
2308.3$-$0211 & 0.2966  &  23  08 22.3  &  $-02$  11 32.1  &  4.45  & $0.27 \pm0.02$ & 7.31 &$ 7.9 \pm  0.7 $&$
0.31 \pm 0.14$ 
& $ 1.20 \pm 0.13$ & $0.88 \pm 0.11$\\
2337.6$+$0016 & 0.2753  &  23  37 37.8  &  $+00$  16 15.5  &  3.82  & $0.27 \pm0.01$ & 7.27 &$ 9.6 \pm  0.3 $&$
0.25 \pm 0.06$ 
& $ 1.00 \pm 0.09$ & $0.79 \pm 0.08$\\
\hline
\hline
  \end{tabular}
  \end{center}
\hspace*{0.3cm}{\footnotesize
}
  }
\end{table*}

\begin{table*} { \begin{center} \footnotesize
  {\renewcommand{\arraystretch}{1.3} \caption[]{ Deduced properties of
  the REFLEX-DXL clusters. Column~(1): cluster name; Col.~(2): central
  electron number density; Col.~(3): central entropy; Cols.~(4,5):
  cooling time and cooling radius; Col.~(6): $r_{500}$; Cols.~(7--9):
  gas mass, total mass, and gas mass fraction
  at $r_{500}$. }
  \label{t:catalog1}}
  \begin{tabular}{lllllllll}
\hline
\hline
RXCJ & $n_{e0}$        & $S_{0} $      & $t_{\rm c}
$ & $r_{\rm cool}$ & $r_{500}$ & $M_{\rm gas, 500}$ & $M_{500}$
& $f_{\rm gas, 500}$ \\
\multicolumn{2}{r}{($10^{-3}$~cm$^{-3}$)} & (keV~cm$^{2}$) &
(Gyr) & (Mpc) &  (Mpc) &($10^{14} M_{\odot}$)& ($10^{14} M_{\odot}$)& \\
\hline
0014.3$-$3022    & $    3.7  \pm     0.2 $ & $  421  \pm    37 $ & $   14.1  \pm     0.6 $ & $    0.00   $ & $    1.24
$ & $    1.0  \pm     0.2 $ & $    7.4  \pm     2.9 $ & $    0.142  \pm     0.024   $\\
0043.4$-$2037    & $    5.8  \pm     0.4 $ & $  293  \pm    17 $ & $   10.2  \pm     0.3 $ & $    0.12   $ & $    1.08  
$ & $    0.6  \pm     0.1 $ & $    4.8  \pm     1.8 $ & $    0.120  \pm     0.023   $\\
0232.2$-$4420    & $   14.0  \pm     0.1 $ & $  279  \pm    10 $ & $   10.3  \pm     0.2 $ & $    0.18   $ & $    1.30  
$ & $    0.9  \pm     0.2 $ & $    8.4  \pm     2.4 $ & $    0.105  \pm     0.023   $\\
0303.7$-$7752    & $    6.0  \pm     0.4 $ & $  311  \pm    24 $ & $   10.3  \pm     0.4 $ & $    0.12   $ & $    1.27  
$ & $    0.8  \pm     0.1 $ & $    7.7  \pm     2.3 $ & $    0.101  \pm     0.022   $\\
0307.0$-$2840    & $   10.2  \pm     0.3 $ & $  265  \pm    16 $ & $   10.5  \pm     0.3 $ & $    0.17   $ & $    1.14  
$ & $    0.6  \pm     0.1 $ & $    5.5  \pm     1.1 $ & $    0.113  \pm     0.029   $\\
0437.1$+$0043    & $   10.1  \pm     0.4 $ & $  223  \pm    14 $ & $   10.2  \pm     0.3 $ & $    0.16   $ & $    1.17  
$ & $    0.5  \pm     0.1 $ & $    6.1  \pm     2.2 $ & $    0.081  \pm     0.020   $\\
0516.7$-$5430    & $    2.9  \pm     0.2 $ & $  403  \pm    24 $ & $   16.7  \pm     0.4 $ & $    0.00   $ & $    1.19  
$ & $    0.8  \pm     0.2 $ & $    6.4  \pm     2.1 $ & $    0.122  \pm     0.029   $\\
0528.9$-$3927    & $   10.6  \pm     0.4 $ & $  280  \pm    16 $ & $   10.3  \pm     0.3 $ & $    0.14   $ & $    1.19  
$ & $    0.9  \pm     0.1 $ & $    6.4  \pm     1.5 $ & $    0.135  \pm     0.028   $\\
0532.9$-$3701    & $   13.3  \pm     0.5 $ & $  337  \pm    23 $ & $   10.4  \pm     0.3 $ & $    0.14   $ & $    1.13  
$ & $    0.6  \pm     0.1 $ & $    5.4  \pm     1.7 $ & $    0.107  \pm     0.024   $\\
0658.5$-$5556    & $    6.1  \pm     0.2 $ & $  354  \pm    16 $ & $   10.1  \pm     0.2 $ & $    0.15   $ & $    1.42  
$ & $    1.8  \pm     0.3 $ & $   11.0  \pm     6.8 $ & $    0.161  \pm     0.018   $\\
1131.9$-$1955    & $    5.8  \pm     0.3 $ & $  307  \pm    18 $ & $   10.0  \pm     0.3 $ & $    0.11   $ & $    1.10  
$ & $    0.8  \pm     0.2 $ & $    5.2  \pm     3.0 $ & $    0.160  \pm     0.030   $\\
2308.3$-$0211    & $    9.2  \pm     0.3 $ & $  296  \pm    22 $ & $   10.1  \pm     0.3 $ & $    0.14   $ & $    1.24  
$ & $    0.7  \pm     0.1 $ & $    7.4  \pm     1.8 $ & $    0.089  \pm     0.022   $\\
2337.6$+$0016    & $    5.7  \pm     0.3 $ & $  372  \pm    21 $ & $   12.3  \pm     0.3 $ & $    0.00   $ & $    1.43  
$ & $    0.8  \pm     0.1 $ & $   10.9  \pm     2.6 $ & $    0.074  \pm     0.016   $\\
\hline
Mean              &  ---                   & ---                 & ---                     & ---           & ---        
  & ---                     & ---                     & $ 0.116 \pm 0.007 $           \\
\hline
\hline
  \end{tabular}
  \end{center}
\hspace*{0.3cm}{\footnotesize
}
  }
\end{table*}

\begin{table*} { \begin{center} \footnotesize
  {\renewcommand{\arraystretch}{1.3} \caption[]{ Parameters of the
  extended-NFW model fits of the REFLEX-DXL clusters. Column~(1):
  cluster name; Cols.~(2,3): characteristic density and scale of the
  halo; Cols.~(4): slope parameter; Col.~(5): $\chi^2$ and its degree
  of freedom (d.o.f.); Col.~(6): concentration parameter at $r_{500}$.}
  \label{t:catalog2}} \begin{tabular}{lllrrr}
\hline
\hline
RXCJ          & $\rho_{\rm s}$            & $r_{\rm s}  $ & $\alpha$&$\chi^2/$d.o.f.&$c_{500}$    \\
              & ($M_{\odot}$~Mpc$^{-3}$)  & (Mpc)         &         &               &             \\
\hline
0014.3$-$3022 & $(2.7\pm 0.3)\times 10^4$ & $0.16\pm0.01$ & $-1.69$ &90.0/157       &$ 7.5\pm1.5 $\\
0043.4$-$2037 & $(1.8\pm 0.2)\times 10^4$ & $0.15\pm0.01$ & $-0.56$ &30.9/115       &$ 7.4\pm1.5 $\\
0232.2$-$4420 & $(1.3\pm 0.7)\times 10^2$ & $1.13\pm0.74$ & $ 1.28$ &65.7/124       &$ 1.2\pm0.4 $\\
0303.7$-$7752 & $(8.8\pm 0.6)\times 10^3$ & $0.22\pm0.01$ & $-0.17$ &18.0/143       &$ 5.7\pm1.2 $\\
0307.0$-$2840 & $(3.7\pm 0.2)\times 10^4$ & $0.11\pm0.01$ & $-0.87$ &3.8/82         &$10.7\pm2.2 $\\
0437.1$+$0043 & $(6.2\pm 0.7)\times 10^3$ & $0.23\pm0.01$ & $ 0.08$ &34.8/97        &$ 5.1\pm1.1 $\\
0516.7$-$5430 & $(2.5\pm 0.2)\times 10^3$ & $0.38\pm0.02$ & $-0.09$ &21.4/131       &$ 3.1\pm0.6 $\\
0528.9$-$3927 & $(4.6\pm 2.6)\times 10^4$ & $0.08\pm0.02$ & $-0.51$ &113.8/56       &$14.6\pm4.4 $\\
0532.9$-$3701 & $(6.9\pm 1.1)\times 10^4$ & $0.80\pm0.01$ & $-0.65$ &39.7/118       &$14.2\pm3.0 $\\
0658.5$-$5556 & $(4.0\pm 0.6)\times 10^3$ & $0.43\pm0.03$ & $-0.27$ &333.6/240      &$ 3.3\pm0.7 $\\
1131.9$-$1955 & $(1.2\pm 0.3)\times 10^6$ & $0.04\pm0.01$ & $-7.71$ &49.9/128       &$30.0\pm6.6 $\\
2011.3$-$5725 & $(4.3\pm 4.1)\times 10^4$ & $0.07\pm0.03$ & $-0.44$ &7.0/8          &$ 4.2\pm0.5 $\\
2308.3$-$0211 & $(2.4\pm 0.3)\times 10^4$ & $0.13\pm0.01$ & $-0.31$ &9.4/58         &$ 9.7\pm2.0 $\\
2337.6$+$0016 & $(6.8\pm 1.9)\times 10^3$ & $0.23\pm0.03$ & $ 0.12$ &82.1/91        &$ 6.3\pm1.5 $\\
\hline
\hline
  \end{tabular}
  \end{center}
\hspace*{0.3cm}{\footnotesize
}
  }
\end{table*}

\begin{table*} { \begin{center} \footnotesize
  {\renewcommand{\arraystretch}{1.3} \caption[]{ Power law,
  $Y=Y_0\;X^{\gamma}$, parameterized scaling relations in this work and
  in literature.}  \label{t:mtx_lite}}
\begin{tabular}{lllll}
\hline
\hline
$Y$ & $X$ & $Y_0$ & $\gamma$ & Reference \\
\hline
$\frac{M_{500}}{M_{\odot}} \; E(z) \; (\Delta_{c,z}/\Delta_{c,0})^{0.5}$ & $\frac{T}{\rm keV}$
   & $10^{13.85}h_{50}^{-1}$ & 1.5 (fixed) & Evrard et al. 96 \\
 & & $10^{13.63^{+0.08}_{-0.07}}h_{50}^{-1}$ & $1.48^{+0.10}_{-0.12}$ & Finoguenov et al. 01b \\
 & & $10^{13.96\pm0.02}h_{70}^{-1}$ & 1.5 (fixed) & Ettori et al. 04 \\
 & & $10^{13.46\pm0.02}h_{70}^{-1}$ & $1.59\pm0.05$ & Popesso et al. 05 \\
 & & $10^{13.57\pm0.02}h_{70}^{-1}$ & $1.49\pm0.15$ & Arnaud et al. 05 \\
 & & $10^{13.80\pm0.04}h_{70}^{-1}$ & 1.5 (fixed) & This work \\
\hline
$\frac{M_{\rm gas,500}}{M_{\odot}} \; E(z) \; (\Delta_{c,z}/\Delta_{c,0})^{0.5}$ & $\frac{T}{\rm keV}$
   & $10^{12.63\pm0.03}h_{50}^{-1}$ & $1.98\pm0.18$ & Mohr et al. 99\\
 & & $10^{12.60}h_{50}^{-1}$ & $1.80\pm0.16$ & Castillo-Morales \& Schindler 03\\
 & & $10^{12.53\pm0.04}h_{70}^{-1}$ & 1.8 (fixed) & This work \\
\hline
$\frac{L_{\rm X}^{\rm 0.1-2.4 keV}}{\rm erg\;s^{-1}}\;E(z)^{-1} \; (\Delta_{c,z}/\Delta_{c,0})^{-0.5}$ & $\frac{T}{\rm keV}$
   & $10^{42.52\pm0.04}h_{100}^{-1}$ & $2.10\pm0.24$ & Markevitch 98 \\
 & & $10^{42.79\pm0.09}h_{50}^{-1}$ & $2.60\pm0.13$ & Reiprich \& B\"ohringer 02 \\
 & & $10^{42.19\pm0.29}h_{50}^{-1}$ & $2.44\pm0.39$ & Ikebe et al. 02 \\
 & & $10^{42.37\pm0.06}h_{70}^{-1}$ & 2.60 (fixed) & This work \\
\hline
$\frac{L_{\rm bol}}{\rm erg\;s^{-1}}\;E(z)^{-1} \; (\Delta_{c,z}/\Delta_{c,0})^{-0.5}$ & $\frac{T}{\rm keV}$
   & $10^{42.43\pm0.04}h_{100}^{-1}$ & $2.64 \pm 0.27$ & Markevitch 98 \\
 & & $10^{42.82\pm0.03}h_{50}^{-1}$ & $2.88 \pm 0.15$ & Arnaud \& Evrard 99 \\
 & & $10^{42.97\pm0.07}h_{50}^{-1}$ & $2.79 \pm 0.08$ & Xue \& Wu 00 \\
 & & $10^{42.85\pm0.09}h_{50}^{-1}$ & $2.98 \pm 0.12$ & Reiprich \& B\"ohringer 02 \\
 & & $10^{42.38\pm0.06}h_{70}^{-1}$ & 2.98 (fixed) & This work \\
\hline
$\frac{L_{\rm X}^{\rm 0.1-2.4 keV}}{\rm erg\;s^{-1}}\;E(z)^{-1} \; (\Delta_{c,z}/\Delta_{c,0})^{-0.5}$ &
$\frac{M_{500}}{M_{\odot}} \; E(z) \; (\Delta_{c,z}/\Delta_{c,0})^{0.5}$ 
   & $10^{22.455\pm1.298}h_{50}^{-1}$ & $1.504\pm0.089$ & Reiprich \& B\"ohringer 02 \\
 & & $10^{25.19\pm0.10}h_{70}^{-1}$ & $1.30\pm0.12$ & Popesso et al. 05 \\
 & & $10^{25.00\pm0.06}h_{70}^{-1}$ & 1.3 (fixed) & This work \\
\hline
$\frac{L_{\rm bol}}{\rm erg\;s^{-1}}\;E(z)^{-1} \; (\Delta_{c,z}/\Delta_{c,0})^{-0.5}$ &
$\frac{M_{500}}{M_{\odot}} \; E(z) \; (\Delta_{c,z}/\Delta_{c,0})^{0.5}$  
   & $10^{25.35\pm0.06}h_{70}^{-1}$ & 1.3 (fixed)       & This work \\
\hline
$f_{\rm gas,500}$ & $\frac{T}{\rm keV}$
  & $0.035 \pm 0.030 h^{-3/2}_{70}$          & $0.5 \pm 0.3$     & This work \\
\hline
\hline
  \end{tabular}
  \end{center}
\hspace*{0.3cm}{}
  }
\end{table*}

\begin{table*} { \begin{center} \footnotesize
  {\renewcommand{\arraystretch}{1.3} \caption[]{Logarithmic intrinsic
  scatter measured for the scaling relations. Column(1): scaling
  relation; Col.(2): variable; Cols.(3,4): logarithmic intrinsic
  scatter and correlative scatter.}  \label{t:scaint}}
\begin{tabular}{llll}
\hline
\hline
Relation & Variable & $\overline{\sigma}_{\rm int}$ & $\overline{\sigma}_{\rm cor}$ \\
\hline
$M_{500}$--$T$         & $\lg(M_{500})$ & 0.17 & 0.20\\
                       & $\lg(T)$ & 0.09 & 0.10\\
$M_{\rm gas,500}$--$T$ & $\lg(M_{\rm gas,500})$ & 0.14 & 0.16\\
                       & $\lg(T)$ & 0.07 & 0.08\\
$L_{X}^{\rm 0.1-2.4 keV}$--$T$ & $\lg(L_{X}^{\rm 0.1-2.4 keV})$ & 0.16 & 0.17\\
                               & $\lg(T)$ & 0.06 & 0.07\\
$L_{X}^{\rm 0.1-2.4 keV}$--$M_{500}$ & $\lg(L_{X}^{\rm 0.1-2.4 keV})$ & 0.19 & 0.20\\
                               & $\lg(M_{500})$ & 0.16 & 0.19\\
$L_{\rm bol}$--$T$             & $\lg(L_{\rm bol}$ & 0.16 & 0.17\\
                               & $\lg(T)$ & 0.06 & 0.07\\
$L_{\rm bol}$--$M_{500}$       & $\lg(L_{\rm bol}$ & 0.18 & 0.19\\
                               & $\lg(M_{500})$ & 0.17 & 0.20\\
\hline
\hline
  \end{tabular}
  \end{center}
\hspace*{0.3cm}{\footnotesize } }
\end{table*}


\clearpage



\clearpage

\clearpage

\clearpage


\end{document}